\documentclass[aps,prb,twocolumn,superscriptaddress,10pt,floatfix]{revtex4-2}
\usepackage{silence}
\usepackage{placeins}
\usepackage{booktabs}
\usepackage[T1]{fontenc}
\usepackage{amsmath,amssymb,amsfonts}
\usepackage{graphicx}
\usepackage{bm}
\usepackage{tikz}

\usepackage{hyperref}
\hypersetup{colorlinks=true,
    linkcolor=blue,
    citecolor=blue,
    urlcolor=blue}

\begin{document}

\title{Rotational symmetry and common-mode phase drift in a
counter-wound S-shaped Aharonov--Bohm interferometer}

\author{Halil Serdar Solak}
\affiliation{Department of Chemical Engineering, Faculty of Chemistry and Metallurgy, Yildiz Technical University, İstanbul, Türkiye}

\author{Afif Sıddıki}
\affiliation{Vocational School, Atlas University, 34408, İstanbul, Türkiye}

\date{\today}

\begin{abstract}
We study common flux offsets in a three-dimensional InAs
Aharonov--Bohm waveguide formed by two quarter-circle bends of
opposite curvature. Each bend contains a through-opening that
separates two conducting arms and encloses an independently
specified flux. For a spin-independent, two-terminal model with
ideal confined fluxes, microreversibility and a rotation exchanging
the complete modules imply
$\mathcal T(\phi_1,\phi_2)=\mathcal T(-\phi_2,-\phi_1)$.
The counter-wound configuration is therefore stationary against
common offsets, while the same-winding configuration is stationary
against differential offsets. This constraint does not require
cancellation of every interference phase, and it extends to a
spatially uniform ambient field perpendicular to the plane of the guide,
even though that field also penetrates the conductor.
Three-dimensional scattering calculations at $R=350\,\mathrm{nm}$ and
$E_F\simeq7.283\,\mathrm{meV}$ retain three open orbital lead modes
and strong reflection at the openings. The finest sampled
counter-wound sweep has a visibility of approximately $3.56\%$; its
grid dependence is dominated by a discretization shift of the single
open mode of each arm, and referring the grids to a common threshold
gives an estimated continuum visibility of 2.6--2.9\%. The response
coefficients vary strongly with energy, the common-mode curvature
changing sign within about $0.1\,\mathrm{meV}$ of $E_F$, while a
$100\,\mathrm{mK}$ electron temperature retains about $98\%$ of the
visibility. Full common-offset
sweeps test the local Taylor estimates of drift tolerance.
For weak, purely common Gaussian phase noise, the leading
transmission variance is quartic in noise amplitude, whereas the
mean shift remains quadratic. Flux imbalance and asymmetric scalar
potentials restore a linear response; symmetric scalar disorder
preserves stationarity without guaranteeing high transmission.
The proposed experimental test requires few-mode coherent
transport, calibrated flux controls, and a magnetic-source design
whose leakage fields and spin-dependent terms are assessed
explicitly.
\end{abstract}

\maketitle

\section{Introduction}
\label{sec:intro}

The Aharonov--Bohm effect relates electron interference to an enclosed
magnetic flux even when the magnetic field vanishes along the accessible
paths~\cite{ModelAB1959}. Its gauge structure and experimental
interpretation have been examined extensively~\cite{Olariu1985,PeshkinTonomura1989},
including electron-interference measurements with a magnetically
shielded flux source~\cite{FeasTonomura1986}. In a conductor attached to
reservoirs, the magnetic phase enters the scattering amplitudes and
therefore the conductance. Early analyses of rings connected to leads
related this response to flux-dependent transmission and
resonances~\cite{Gefen1984,Buttiker1984}. Oscillations with the $h/e$
flux period were subsequently observed in normal-metal
rings~\cite{Webb1985}; the associated coherence and transport phenomena
are reviewed in Ref.~\cite{Washburn1986}.

Semiconductor structures extended these experiments to geometries with
electrostatic control. Aharonov--Bohm oscillations were observed in
GaAs/AlGaAs devices~\cite{Timp1987}, and interferometry was used to
resolve a coherent contribution to transmission through a quantum
dot~\cite{Yacoby1995}. Conductance quantization in narrow point
contacts established the role of individual propagating
channels~\cite{vanWees1988,Wharam1988}. More recent work has explored
electronic flying-qubit interferometers~\cite{Bautze2014} and
sound-driven transfer between coupled quantum rails~\cite{Takada2019}.
Reviews of single-electron control and semiconductor flying qubits
place these approaches within the broader effort to prepare, route,
and detect propagating quantum states~\cite{Bauerle2018,Edlbauer2022}.
These developments motivate careful control of an interferometer's
response, although stable total conductance and high single-electron
state fidelity remain different observables.

The same phase sensitivity that produces interference also makes its
measurement susceptible to fluctuations. Coupling to an environment
can generate phase uncertainty and suppress interference~\cite{Stern1990},
while low-frequency noise is a persistent limitation in solid-state
quantum devices~\cite{paladino2014}. Several approaches address this
sensitivity. Operating at a stationary point can reduce the leading
response to a fluctuating control parameter, as illustrated by the
optimal operating point of the quantronium~\cite{Vion2002}.
Dynamical decoupling instead uses time-dependent control to suppress
coupling to an environment~\cite{Viola1999}. Here we examine a static
scattering structure in which a spatial symmetry fixes a direction of
first-order insensitivity in the two-flux parameter space.

Symmetry constraints of this kind are familiar in the single-flux case.
For a two-terminal conductor, unitarity together with time-reversal
symmetry forces the conductance to be an even function of the enclosed
flux, so that the fitted transmission phase is locked to $0$ or $\pi$.
This ``phase rigidity'' underlies the interpretation of transmission-phase
measurements in ring interferometers and of its breakdown in open or
multi-terminal geometries~\cite{LevyYeyati1995,Aharony2002,Aharony2003}.
The constraint examined here is the two-flux extension of that statement:
with two independently controlled fluxes, evenness in a single variable is
replaced by separate evenness in the common and differential coordinates,
and the additional ingredient is a spatial operation exchanging the two
modules and the two contacts.

Opposite fluxes alone do not determine that response. Two
series-connected Aharonov--Bohm interferometers with zero total flux
have already been studied as a quantum-transport problem~\cite{Wang2007}.
Scattering by two opposite flux vortices has also been analyzed in
free space~\cite{SymmetryBogomolny2010}, and the general two-vortex
problem admits an integrable formulation~\cite{SymmetryBogomolny2016}.
For an interference contour winding equally around two opposite
fluxes, the magnetic contribution cancels. Other winding combinations
retain a phase, so vanishing total flux does not imply a flux-independent
transmission. Our question concerns the additional constraint imposed
when a rotation exchanges the two openings and the contacts of a
confined, two-terminal conductor.

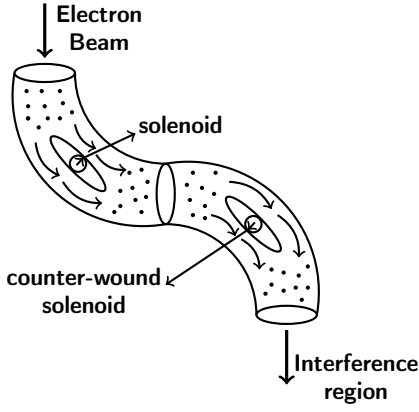
\begin{figure}[t]
    \centering
    \begin{tikzpicture} [scale=0.4]

%\draw[help lines] grid (18,15);

\draw[thick] (4,12) ellipse (1cm and 0.3cm);
\draw[thick] (8,8) ellipse (0.3cm and 1cm);
\draw[thick] (12,4) ellipse (1cm and 0.3cm);

\draw[thick] (3,12) arc (170:280:4.316cm);
\draw[thick] (5,12) arc (180:270:3cm);

\draw[thick] (13,4) arc (-10:100:4.316cm);
\draw[thick] (11,4) arc (0:90:3cm);

\draw[thick, rotate around={-45:(5.1,9)}] (5.1,9) ellipse (1.25cm and 0.3cm);
\draw[thick, rotate around={-45:(10.9,7)}] (10.9,7) ellipse (1.25cm and 0.3cm);

\draw[thick] (5.1,9) circle (0.27cm and 0.27cm);
\draw[thick] (10.9,7) circle (0.27cm and 0.27cm);

\draw[very thick, ->] (4,14.3) -- (4,12.5);
\draw[very thick, ->] (12,3.5) -- (12,1.7);

\draw[thick, ->] (5.05,8.95) -- (5.28,9.2);
\draw[thick, ->] (10.95,7.06) -- (10.72,6.81);

\node[font=\sffamily\small\bfseries, align=center] at (5.8,13.5) {Electron \\ Beam};
\node[font=\sffamily\small\bfseries, align=center] at (14.3,1.9) {Interference \\ region};

\draw[thick, ->] (3.7,9.7) to[bend right] (4.5,8.6);
\draw[thick, ->] (4.6,8.4) to[bend right] (5.9,7.7);

\draw[thick, ->] (5,10.3) to[bend right] (5.7,9.5);
\draw[thick, ->] (5.7,9.3) to[bend right] (6.7,8.8);

\fill (3.4,11.4) circle (2pt);
\fill (3.45,10.9) circle (2pt);
\fill (3.9,11.4) circle (2pt);
\fill (4,10.9) circle (2pt);
\fill (3.45,10.5) circle (2pt);
\fill (3.45,10.5) circle (2pt);
\fill (4.1,10.5) circle (2pt);
\fill (3.8,10.2) circle (2pt);
\fill (4.5,11.4) circle (2pt);
\fill (4.6,11) circle (2pt);
\fill (4.5,10.5) circle (2pt);
\fill (4.9,10.7) circle (2pt);

\draw[thick, ->] (5.3,9.2) -- (7,10);
\node[font=\sffamily\small\bfseries] at (8.5,10.3) {solenoid};

\fill (7,7.8) circle (2pt);
\fill (6.9,8.2) circle (2pt);
\fill (6.8,8.7) circle (2pt);
\fill (7.2,7.4) circle (2pt);
\fill (7.35,7.9) circle (2pt);
\fill (7.4,8.4) circle (2pt);
\fill (7.2,8.8) circle (2pt);
\fill (6.5,7.6) circle (2pt);
\fill (6.3,7.4) circle (2pt);
\fill (6.5,8.1) circle (2pt);

\fill (9.2,8.2) circle (2pt);
\fill (8.9,8.7) circle (2pt);
\fill (8.8,7.8) circle (2pt);
\fill (8.9,7.3) circle (2pt);
\fill (9.6,8) circle (2pt);
\fill (9.3,7.6) circle (2pt);
\fill (9.3,7.2) circle (2pt);
\fill (9.65,8.7) circle (2pt);
\fill (8.7,8.3) circle (2pt);

\draw[thick, ->] (10.7,6.8) -- (8,5);
\node[font=\sffamily\small\bfseries, align=center] at (5.3,4.8) {counter-wound \\ solenoid};

\draw[thick, ->] (9.5,7.3) to[bend left] (10.3,6.6);
\draw[thick, ->] (10.5,6.5) to[bend left] (11.1,5.5);

\draw[thick, ->] (10.1,8.4) to[bend left] (11.5,7.5);
\draw[thick, ->] (11.7,7.3) to[bend left] (12.4,5.9);

\fill (11.5,5) circle (2pt);
\fill (11.3,4.7) circle (2pt);
\fill (11.9,4.8) circle (2pt);
\fill (11.9,5.4) circle (2pt);
\fill (12.4,5.3) circle (2pt);
\fill (12.3,4.9) circle (2pt);
\fill (12.7,4.9) circle (2pt);
\fill (12.5,4.5) circle (2pt);
\fill (12.7,5.6) circle (2pt);
\fill (11.5,5.5) circle (2pt);

\end{tikzpicture}
    \caption{Schematic projection of the three-dimensional double-bend
    waveguide. Two quarter-circle reference arcs of opposite curvature
    are joined tangentially. The tube has an elliptical transverse
    cross section, and a through-opening in each bend creates two
    lateral conducting arms that rejoin before the next split.
    Solenoids occupy the openings without contacting the conductor;
    their axes are normal to the centerline plane. Both signed fluxes
    are referred to the same normal direction, with
    $\Phi_1=-\Phi_2$ for counter-wound operation. Magnetic fields are
    assumed to be confined to the excluded regions. Dots and arrows
    indicate propagation schematically, rather than calculated particle
    trajectories. Opposite fluxes do not imply cancellation of every
    interference phase; the symmetry result concerns the response of
    the total transmission to a common flux offset. Not to scale.}
    \label{fig:schematic}
\end{figure}

The proposed structure is shown schematically in
Fig.~\ref{fig:schematic}. Its reference centerline consists of two
quarter-circle bends of equal radius and opposite curvature. This
centerline is planar, while the conducting waveguide is three dimensional
and has an elliptical transverse cross section. A through-opening in
each bend separates the conductor into two lateral arms that rejoin
before the next split. No conducting bridge connects the arms above
or below an opening. A solenoid occupies each opening without
contacting the conductor. The openings define the branching geometry
and remain present when the magnetic sources are removed.

We use InAs as the material setting for the model. Selective-area
growth has produced patterned InAs channels and Aharonov--Bohm loop
devices exhibiting phase-coherent transport~\cite{FeasLee2019}.
Optimization of InAs/InGaAs structures has also improved their measured
field-effect mobility~\cite{FeasBeznasyuk2022}. These results support
examining an InAs implementation, while leaving the confinement,
elastic scattering, and magnetic-source requirements of this particular
design to be established. A spin-independent effective-mass Hamiltonian
with ideal confined fluxes defines the reference calculation.

The central identity combines two-terminal
microreversibility~\cite{ModelButtiker1986} with the rotation exchanging
the complete modules and their leads. For signed phases
$\phi_j=e\Phi_j/\hbar$, define
$C=(\phi_1+\phi_2)/2$ and $D=(\phi_1-\phi_2)/2$.
The balanced orbital model has a total transmission that is separately
even in $C$ and $D$. Counter-wound operation, with opposite signed
fluxes, lies on $C=0$ and is stationary against common offsets wherever
the derivative exists. Same-winding operation lies on $D=0$ and is
stationary against differential offsets. The two assignments protect
different directions in parameter space; their usefulness depends on
the perturbations present in a sample. The same argument applies to a
spatially uniform drift of the ambient field perpendicular to the plane
of the guide, including its part inside
the conductor, which therefore lies in the protected direction for
counter-wound operation, whereas a common fluctuation of the source
currents in oppositely wound coils generally does not. In this respect
the counter-wound assignment resembles a gradiometric superconducting
loop, in which a figure-eight winding makes the net flux of a uniform
field vanish~\cite{clarke2004squid}. Here the protection is a property
of the total transmission, not of individual interference contours, and
it holds to first order; Sec.~\ref{sec:discussion} compares the two
cases.

A three-dimensional scattering calculation evaluates the response while
retaining transverse confinement, reflection, and recombination at the
openings. Mesh comparisons separate the exact transmission identities
from the numerical accuracy of modulation amplitudes and sensitivity
coefficients. The local analysis also distinguishes fluctuations of
the transmitted signal from a change in its mean: under weak, purely
common Gaussian noise, stationarity suppresses the leading variance,
while the mean shift remains quadratic. Controlled flux and
electrostatic imbalances, together with a spatially correlated scalar
potential, identify how the linear response reappears when the modules
are no longer balanced. Energy-resolved calculations show how strongly
the magnitudes left free by the symmetry depend on energy, trace the
grid dependence of the modulation to a discretization shift of the arm
threshold, and give the contrast at finite temperature.

Section~\ref{sec:model} specifies the geometry and transport assumptions.
Section~\ref{sec:phases} distinguishes magnetic winding phases from
propagation through the bends, and Sec.~\ref{sec:symmetry} derives the
transmission identities. Sections~\ref{sec:decoherence_analysis}
and~\ref{sec:robustness} present the flux response and its sensitivity to
imbalance and disorder. Section~\ref{sec:feasibility} examines the material,
magnetic-source, and measurement requirements. Sections~\ref{sec:discussion}
and~\ref{sec:scope} discuss the interpretation and limitations, followed
by the conclusions in Sec.~\ref{sec:conclusion}.

\section{System model and assumptions}
\label{sec:model}

\subsection{Waveguide geometry}

We consider an InAs waveguide formed by two quarter-circle bends of equal
radius and opposite curvature. The reference centerline lies in the
$xy$ plane, while the conducting region has finite width and height.
In each bend, a through-opening separates the guide into two lateral
arms, which rejoin before the next module. A solenoid occupies each
opening without contacting either arm. The openings and arms are part
of the waveguide geometry and remain present when the solenoids are removed.

Place the junction at the origin. With $\ell_b=\pi R/2$,
$\theta=s/R$, and $s$ the reference arclength, the centerline is
\begin{equation}
\mathbf r_0(s)=R
\begin{cases}
(\sin\theta,\,1-\cos\theta,\,0),
    &-\ell_b\leq s\leq0,\\[2pt]
(\sin\theta,\,\cos\theta-1,\,0),
    &0\leq s\leq\ell_b.
\end{cases}
\label{eq:model_centerline}
\end{equation}
The tangent is continuous at the junction. Define
$\mathbf t=d\mathbf r_0/ds$ and
$\mathbf n=\hat{\mathbf z}\times\mathbf t$.
Before removing the openings, points in the bent tube have the form
\begin{equation}
\begin{gathered}
\mathbf r=\mathbf r_0(s)+u\mathbf n(s)+v\hat{\mathbf z},
\qquad |s|\leq\ell_b,\\
\frac{u^2}{a_h^2}+\frac{v^2}{a_v^2}<1.
\end{gathered}
\label{eq:model_tube}
\end{equation}
Here $a_h$ and $a_v$ are the horizontal and vertical semiaxes of
the transverse ellipse. We take
\begin{equation}
R=350\,\mathrm{nm},\qquad
a_h=80\,\mathrm{nm},\qquad a_v=60\,\mathrm{nm}.
\label{eq:model_dimensions}
\end{equation}
Thus the full width and height are $160$ and $120\,\mathrm{nm}$.
The flattening is in the transverse cross section; the reference bends
remain quarter-circles. Identical straight leads extend tangentially
from the two ends.

The reference length is
$L_{\mathrm{ref}}=\pi R\simeq1.100\,\mu\mathrm{m}$, excluding the leads.
For a curve at a fixed horizontal offset $u$ in the unperforated tube,
the lengths $L_1$ and $L_2$ of its segments in the first and second
bends satisfy
\begin{equation}
\begin{gathered}
L_1(u)=\frac{\pi}{2}(R-u),\qquad
L_2(u)=\frac{\pi}{2}(R+u),\\
L_1(u)+L_2(u)=\pi R.
\end{gathered}
\label{eq:pathbalance}
\end{equation}
This is a geometric reference identity. After the openings are removed,
some such curves intersect the excluded region, including the centerline
itself. Equation~\eqref{eq:pathbalance} does not assign a trajectory to
an electron or establish equality of all transmitted propagation phases.

\subsection{Openings and confined fluxes}

The openings are centered at
$\mathbf q_1=\mathbf r_0(-\ell_b/2)$ and
$\mathbf q_2=\mathbf r_0(\ell_b/2)=-\mathbf q_1$.
At each center, introduce
$\xi_j=(\mathbf r-\mathbf q_j)\cdot\mathbf t_j$ and
$\eta_j=(\mathbf r-\mathbf q_j)\cdot\mathbf n_j$, with the frame
evaluated at the corresponding midpoint. The conducting domain excludes
\begin{equation}
\mathcal H_j=\left\{\mathbf r:
\frac{\xi_j^2}{b_o^2}+\frac{\eta_j^2}{a_o^2}\leq1\right\},
\qquad j=1,2.
\label{eq:model_openings}
\end{equation}
Each opening extends through the full vertical thickness of the tube.
The two arms connect around its longitudinal ends, with no conducting
bridge above or below it. The long axis of its elliptical footprint
follows the local tangent at the bend midpoint. We adopt
\begin{equation}
a_o=17.5\,\mathrm{nm},\qquad
\frac{b_o}{a_o}=\frac{1.25}{0.30},\qquad
b_o\simeq72.9\,\mathrm{nm}.
\label{eq:model_opening_dimensions}
\end{equation}
The aspect ratio is a design choice of the reference geometry.
The opening's short diameter is therefore $35\,\mathrm{nm}$.
The magnetic core and the complete solenoid assembly have separate
dimensions and must fit inside this opening.

The solenoid axes are parallel to $\hat{\mathbf z}$.
Their signed fluxes $\Phi_1$ and $\Phi_2$, both referred to
$+\hat{\mathbf z}$, are independent control parameters. We impose
\begin{equation}
\begin{gathered}
\boldsymbol\nabla\times\mathbf A=0
\quad\text{in the conducting domain},\\
\oint_{C_j}\mathbf A\cdot d\boldsymbol\ell=\Phi_j,\\
\phi_j=\frac{e\Phi_j}{\hbar}
=\frac{2\pi\Phi_j}{\Phi_0},\qquad \Phi_0=\frac{h}{e}.
\end{gathered}
\label{eq:model_flux}
\end{equation}
Here $e>0$ and $C_j$ winds counterclockwise about opening $j$ alone,
viewed from $+\hat{\mathbf z}$. For an electron of charge $-e$,
the magnetic phase around this contour is $-\phi_j$.
We call $(\phi_1,\phi_2)=(\phi,-\phi)$ counter-wound and
$(\phi_1,\phi_2)=(\phi,\phi)$ same-winding, keeping the geometry fixed.

For a closed interference contour with winding numbers $w_1$ and $w_2$,
the magnetic phase is $-(w_1\phi_1+w_2\phi_2)$. Opposite fluxes cancel
this contribution when $w_1=w_2$; they do not remove the flux dependence
of every scattering amplitude. Perfect confinement is an assumption
of the magnetic model. A finite solenoid with an iron core requires a
separate assessment of its stray field.

\subsection{Electronic Hamiltonian}

We describe conduction electrons in InAs within a parabolic, isotropic
effective-mass model, using $m^*=0.023m_e$ as the reference
parameter~\cite{IoffeInAsParameters}. This value lies at the lower end
of the reported range of $\Gamma$-valley masses, $0.023$--$0.030\,m_e$;
the recommended low-temperature value of a standard compilation is
$0.026\,m_e$~\cite{Vurgaftman2001}. We retain $m^*=0.023m_e$ as a model
parameter. It enters the calculated response strongly: at fixed $E_F$,
replacing $0.023m_e$ by $0.026m_e$ shifts the accumulated longitudinal
phase of the lowest lead mode over $L_{\rm ref}$ by about $5\,$rad. The
modulation amplitudes and sensitivity coefficients reported below are
therefore specific to the stated mass, whereas the symmetry identities
of Sec.~\ref{sec:symmetry} do not involve $m^*$.
In the accessible three-dimensional domain $\Omega$, including the leads,
\begin{equation}
H=\frac{[-i\hbar\boldsymbol\nabla+e\mathbf A(\mathbf r)]^2}{2m^*}
+V(\mathbf r).
\label{eq:model_hamiltonian}
\end{equation}
Dirichlet conditions apply at the outer walls and the boundaries of the
openings. The clean reference has $V=0$ in $\Omega$. Curvature and
branching enter through the domain itself; no additional one-dimensional
curvature potential is added to Eq.~\eqref{eq:model_hamiltonian}.

The reference Fermi energy is $E_F=7.283\,\mathrm{meV}$, measured from
the interior conduction-band edge. It is a chosen operating parameter,
to be set physically by carrier density and electrostatic control.
In either straight lead, the transverse thresholds follow from
\begin{equation}
-\frac{\hbar^2}{2m^*}
(\partial_u^2+\partial_v^2)\chi_n=E_n^\perp\chi_n,
\qquad \chi_n\big|_{\partial D}=0,
\label{eq:model_lead_modes}
\end{equation}
where $D$ is the transverse ellipse. A mode propagates when
$E_F>E_n^\perp$, with
$k_n=\sqrt{2m^*(E_F-E_n^\perp)}/\hbar$.
This eigenproblem retains confinement in both transverse directions.
The local spectrum in either narrowed arm must be evaluated separately;
it is given in Sec.~\ref{sec:cw_sweep}.
We impose no single-mode condition, equal splitting, or reflectionless
recombination.

The reference Hamiltonian omits spin--orbit and Zeeman coupling,
inelastic processes, and self-consistent interaction corrections.
Spin supplies two identical copies of the orbital problem. These choices
define the reference calculation; the omitted terms must be assessed
when comparing with a fabricated InAs device.

\subsection{Balance and transport assumptions}

The rotation
\begin{equation}
C_2:(x,y,z)\mapsto(-x,-y,z)
\label{eq:model_rotation}
\end{equation}
exchanges the bends, openings, and leads. It can be generated by
successive rotations by $\pi$ about $x$ and $y$ and acts on the tube
coordinates as $(s,u,v)\mapsto(-s,-u,v)$.
Geometric balance means that the entire conducting domain is invariant
under this operation. Any scalar potential in the balanced model must
also satisfy $V(C_2\mathbf r)=V(\mathbf r)$.
A proper rotation about $z$ exchanges the fluxes without reversing
their signs; the counter-wound reversal is a separate operation.

Transport is elastic and phase coherent, between two ideal nonmagnetic
reservoirs in linear response. This reservoir description follows the standard mesoscopic scattering
framework~\cite{Imry2002,Beenakker1991,Datta1995}. Let $t$ be the transmission matrix for
one spin copy and $\mathcal T=\operatorname{Tr}(t^\dagger t)$.
The conductance follows the Landauer--B\"uttiker scattering
formulation~\cite{Landauer1957,ModelButtiker1986,Khomyakov2005Transport}
\begin{equation}
G=\frac{2e^2}{h}\int dE\,
\left(-\frac{\partial f(E;E_F,T_{\rm el})}{\partial E}\right)
\mathcal T(E;\phi_1,\phi_2),
\label{eq:model_conductance}
\end{equation}
where $f$ is the reservoir Fermi function. At zero temperature,
$G=(2e^2/h)\mathcal T(E_F;\phi_1,\phi_2)$.
The two-terminal resistance is $R_{\rm 2t}=1/G$, including the quantum
contact contribution. For a specified flux sweep, we define the
conductance visibility by
\begin{equation}
\mathcal V=\frac{G_{\max}-G_{\min}}{G_{\max}+G_{\min}}.
\label{eq:model_visibility}
\end{equation}
Comparison with a physical sample requires independently established
elastic mean free paths and phase-coherence lengths. Their adequacy
must be checked against the device scale and scattering dwell times;
it does not follow from the choice of InAs.

\section{Phase structure}
\label{sec:phases}

The two openings make the conducting domain multiply connected.
Their fluxes affect interference even when the magnetic field vanishes
throughout the accessible region~\cite{ModelAB1959}.
The resulting transmission is nevertheless a property of the complete
scattering problem. In particular, the two arms of the first bend rejoin
before the wave divides again in the second bend. There is no conserved
arm label connecting the entrance to the exit.

\subsection{Magnetic phase and winding}

Consider two paths with the same endpoints and orient the closed contour
$\mathcal C$ along the first path and back along the second.
With the electron-charge convention of Sec.~\ref{sec:model},
their magnetic phase difference is
\begin{equation}
\Delta\varphi_{\mathcal C}^{\rm AB}
=-\frac{e}{\hbar}\oint_{\mathcal C}
\mathbf A\cdot d\boldsymbol\ell
=-w_1\phi_1-w_2\phi_2.
\label{eq:AB_contour}
\end{equation}
Here $w_j$ is the signed winding number about opening $j$, and
$\phi_j=e\Phi_j/\hbar$ is the flux parameter defined in
Eq.~\eqref{eq:model_flux}. The closed-contour phase is gauge invariant
modulo $2\pi$. An absolute magnetic phase assigned to an open path
does not have the same status.

For the counter-wound assignment, $\phi_1=\phi$ and $\phi_2=-\phi$,
\begin{equation}
\Delta\varphi_{\mathcal C}^{\rm AB}=-(w_1-w_2)\phi.
\label{eq:ABcancel}
\end{equation}
Thus contours with equal winding numbers have zero magnetic phase.
A contour enclosing only one opening generally retains a nonzero phase.
Recombination, reflection, and repeated circulation allow contributions
with different winding numbers. The equality $\Phi_1+\Phi_2=0$ therefore
does not imply a flux-independent transmission or a universal vanishing
phase difference at the output.

The flux dependence remains periodic:
\begin{equation}
\begin{split}
\mathcal T(E;\phi_1+2\pi,\phi_2)&=\mathcal T(E;\phi_1,\phi_2),\\
\mathcal T(E;\phi_1,\phi_2+2\pi)&=\mathcal T(E;\phi_1,\phi_2).
\end{split}
\label{eq:phase_periodicity}
\end{equation}
These identities concern ideal confined fluxes in a fixed domain.
Changing a physical solenoid current or magnetization realizes such
a sweep only to the extent that it leaves the field and other
Hamiltonian parameters in the conducting region unchanged.

\subsection{Propagation through the bends}

Equation~\eqref{eq:pathbalance} balances the lengths of fixed-offset
reference curves in the unperforated tube. More general curves have
a different length. For a path segment expressible as
$\mathbf r(s)=\mathbf r_0(s)+u(s)\mathbf n(s)+v(s)\hat{\mathbf z}$,
\begin{equation}
\mathcal L[u,v]=\int ds\,
\sqrt{[1-\kappa(s)u(s)]^2+[u'(s)]^2+[v'(s)]^2},
\label{eq:general_path_length}
\end{equation}
where the signed curvature is $+1/R$ in the first bend and $-1/R$
in the second. Avoiding an opening generally requires the transverse
coordinates to vary. Paths with backtracking must be treated in
separate segments.

Propagation phases also depend on transverse confinement.
Where a local adiabatic mode description is valid, a propagating mode
has the approximate phase
\begin{equation}
\begin{gathered}
\gamma_n^{\rm prop}=\int k_n(s)\,ds,\\
k_n(s)=\frac{\sqrt{2m^*[E-E_n^\perp(s)-V_n(s)]}}{\hbar},
\end{gathered}
\label{eq:local_propagation_phase}
\end{equation}
with $V_n$ the projected scalar potential.
Different modes need not have the same $k_n$, even for equal
geometric lengths. Near a splitting or recombination region,
intermode coupling and reflection also contribute to the scattering
amplitudes. The full three-dimensional model does not assume the
adiabatic separation used in Eq.~\eqref{eq:local_propagation_phase},
nor does it replace locally evanescent contributions by propagating
rays.

\subsection{Role of curvature}

The planar reference curve admits the continuous frame
$(\mathbf t,\mathbf n,\hat{\mathbf z})$ used in Sec.~\ref{sec:model}.
Its transverse frame has no rotation about the tangent:
$\mathbf n'(s)\cdot\hat{\mathbf z}=0$.
This removes a torsional frame contribution in an untwisted
guide. It does not remove the dependence of the scattering matrix
on curvature and confinement, or establish a zero Berry phase for
every possible spin or orbital extension. Geometric phases arise in adiabatic quantum evolution~\cite{Berry1984},
and constrained-system reductions can produce effective gauge fields
as well as scalar corrections~\cite{Mitchell2001}. The absence of a
torsional frame term here therefore addresses a specific geometric
contribution, rather than every possible geometric phase.

For comparison, reduction to an ideal thin curved wire gives the
curvature potential~\cite{ModelDaCosta1981}
\begin{equation}
V_g=-\frac{\hbar^2\kappa^2}{8m^*}.
\label{eq:thin_wire_curvature}
\end{equation}
At $R=350\,\mathrm{nm}$ and $m^*=0.023m_e$, this is approximately
$-3.38\,\mu\mathrm{eV}$. It is a thin-wire estimate, not an additional
term in the three-dimensional Hamiltonian. Its small magnitude
relative to $E_F$ does not bound reflection or mode conversion at
the openings. Those effects are determined by solving the wave
equation in the actual domain. The existence of curvature-induced bound states in Dirichlet waveguides
illustrates that shaping the domain alone can modify its
spectrum~\cite{Exner1989,Goldstone1992}. Those results motivate retaining
the actual confinement; they do not establish a bound state in the
present perforated geometry.

The useful balance of the device is consequently a symmetry of its
Hamiltonian and terminals. The next section derives its consequence
for total transmission without assuming cancellation of every
propagation phase.

\section{Transmission symmetry}
\label{sec:symmetry}

We use the spin-independent Hamiltonian and the two identical leads
defined in Sec.~\ref{sec:model}. The fluxes are the only
time-reversal-odd parameters, and the conducting domain and scalar
potential are invariant under $C_2$.
Let $\mathcal T_{R\leftarrow L}$ denote the total transmission from
the left reservoir to the right, summed over all propagating orbital
channels at energy $E$. The identities below concern this sum;
they need not hold for an arbitrarily selected incident mode.

\subsection{Reciprocity and spatial rotation}

For a unitary two-terminal scattering matrix,
$\mathcal T_{R\leftarrow L}=\mathcal T_{L\leftarrow R}$ at fixed
parameters. Microreversibility relates transmission in one direction
to transmission in the opposite direction with both fluxes
reversed~\cite{ModelButtiker1986}. Together, these give
\begin{equation}
\mathcal T(E;\phi_1,\phi_2)
=\mathcal T(E;-\phi_1,-\phi_2).
\label{eq:TRS}
\end{equation}

The proper rotation $C_2$ exchanges the openings and the leads.
An axial flux retains its sign under this rotation, so the transformed
flux assignment is $(\phi_2,\phi_1)$.
Using equality of the two directional total transmissions once more,
\begin{equation}
\mathcal T(E;\phi_1,\phi_2)
=\mathcal T(E;\phi_2,\phi_1).
\label{eq:C2}
\end{equation}
This step requires symmetry of the complete scattering structure,
including confinement, scalar potentials, and contacts.
It does not require identifying two particular electron paths.

Combining Eqs.~\eqref{eq:TRS} and \eqref{eq:C2} yields
\begin{equation}
\mathcal T(E;\phi_1,\phi_2)
=\mathcal T(E;-\phi_2,-\phi_1).
\label{eq:composite}
\end{equation}
The identity holds at each energy within the stated model, independent
of the number of open channels and the amount of elastic reflection.

The derivation uses $C_2$ only through its action on the openings, the
leads, and the flux labels, and the same conclusion follows from any
spatial operation with that action. A reflection through a vertical
plane, one containing $\hat{\mathbf z}$, that exchanges the two modules
reverses the circulation sense and therefore the sign of each axial flux; it delivers Eq.~\eqref{eq:composite} directly, and
Eq.~\eqref{eq:C2} then follows with Eq.~\eqref{eq:TRS}. The identities
are therefore not specific to opposite curvature: a mirror-symmetric
guide whose two bends have the same sense, or a straight guide carrying
two mirror-related openings, is likewise stationary against common
offsets. What the balance requires is that a single operation map the
complete scattering structure---domain, scalar potential, and both
contacts---onto itself while exchanging the modules. The counter-wound
geometry is adopted here for its layout: its two bends cancel in net
bend, so that the entrance and exit tangents of
Eq.~\eqref{eq:model_centerline} are parallel and the leads are displaced
laterally by $2R$, whereas a same-curvature pair of equal radius
produces antiparallel leads.

A related combination, flux reversal together with a rotation that
exchanges the two flux lines, was noted for free-space scattering on
two Aharonov--Bohm vortices carrying opposite
fluxes~\cite{SymmetryBogomolny2010}; there it relates the scattering
amplitude at flux $-\alpha$ to its value at $+\alpha$ with the incident
and scattered angles rotated by $\pi$. That configuration is described
by a single flux magnitude, so the coordinates of
Eq.~\eqref{eq:total_flux_coordinates} do not arise. The two-vortex
problem with two independent fluxes has since been solved in closed
form~\cite{SymmetryBogomolny2016}, without examining parity in the
flux plane.

\subsection{Common and differential coordinates}

Define the total flux coordinates
\begin{equation}
\begin{gathered}
C=\frac{\phi_1+\phi_2}{2},\qquad
D=\frac{\phi_1-\phi_2}{2},\\
F_E(C,D)=\mathcal T(E;C+D,C-D).
\end{gathered}
\label{eq:total_flux_coordinates}
\end{equation}
Rotation makes $F_E$ even in $D$, while the composite identity makes
it even in $C$:
\begin{equation}
F_E(C,D)=F_E(C,-D)=F_E(-C,D).
\label{eq:coordinate_parities}
\end{equation}
Total coordinates $C,D$ must be distinguished from the perturbations
$c,d$ measured relative to an operating point.

For a counter-wound operating point $(\phi,-\phi)$, write
\begin{equation}
\begin{gathered}
\phi_1=\phi+c+d,\qquad \phi_2=-\phi+c-d,\\
\mathcal T_{\rm CW}(c,d)=F_E(c,\phi+d).
\end{gathered}
\label{eq:CW_coordinates}
\end{equation}
Equation~\eqref{eq:coordinate_parities} then gives
\begin{equation}
\mathcal T_{\rm CW}(c,d)=\mathcal T_{\rm CW}(-c,d)
\quad\text{for every }d.
\label{eq:CWduality}
\end{equation}
In particular,
\begin{equation}
\mathcal T_{\rm CW}(\delta,0)=\mathcal T_{\rm CW}(-\delta,0),
\qquad
\left.\partial_c\mathcal T_{\rm CW}(c,d)\right|_{c=0}=0,
\label{eq:evenness}
\end{equation}
where the derivative exists. A differential shift moves the device
along the protected line $C=0$; it does not remove the common-mode
stationarity.

For the same-winding reference,
\begin{equation}
\begin{gathered}
\phi_1=\phi+c+d,\qquad \phi_2=\phi+c-d,\\
\mathcal T_{\rm SW}(c,d)=F_E(\phi+c,d)
=\mathcal T_{\rm SW}(c,-d).
\end{gathered}
\label{eq:SWduality}
\end{equation}
This configuration is stationary against differential offsets.
The two flux assignments therefore protect different directions in
parameter space. Neither relation orders their transmissions or
establishes a universal advantage for one assignment.
At $\phi=m\pi$, with integer $m$, the counter-wound and same-winding
assignments are equivalent modulo the flux periods in
Eq.~\eqref{eq:phase_periodicity}.

\subsection{What first-order stationarity implies}

At a regular counter-wound operating point and fixed $d=0$,
\begin{equation}
\mathcal T_{\rm CW}(c,0)
=\mathcal T_0+\frac{K_c}{2}c^2+O(c^4),
\qquad K_c=\partial_C^2F_E(0,\phi).
\label{eq:CW_expansion}
\end{equation}
Symmetry fixes the absence of odd powers, not the sign or magnitude
of $K_c$. The stationary point may be a maximum or a minimum.
A finite drift tolerance must therefore be obtained from the full
response curve, with a local Taylor estimate identified as such.

For a purely common, quasistatic Gaussian offset of variance
$\sigma_c^2$, Eq.~\eqref{eq:CW_expansion} gives
\begin{equation}
\begin{split}
\langle\mathcal T_{\rm CW}\rangle-\mathcal T_0
 &=\frac{K_c}{2}\sigma_c^2+O(\sigma_c^4),\\
\operatorname{Var}(\mathcal T_{\rm CW})
 &=\frac{K_c^2}{2}\sigma_c^4+O(\sigma_c^6).
\end{split}
\label{eq:CW_noise_moments}
\end{equation}
A reference with a nonzero common-mode slope $A_c$ instead has
$\operatorname{Var}(\mathcal T)=A_c^2\sigma_c^2+O(\sigma_c^4)$.
Thus suppression of small output fluctuations and suppression of
the mean shift are different statements. Noise with a differential
component is not covered by the pure-common-noise variance law.

The transmission identities also survive the thermal average in
Eq.~\eqref{eq:model_conductance}, provided the reservoir distribution
is unchanged by the flux perturbation. They do not determine how
thermal averaging changes the numerical contrast; this is evaluated in
Sec.~\ref{sec:temperature}.

\subsection{Scope of the protecting symmetry}

Static scalar disorder can be separated as
\begin{equation}
V_\pm(\mathbf r)=\frac{V(\mathbf r)\pm V(C_2\mathbf r)}{2}.
\label{eq:symmetry_potential_parts}
\end{equation}
For $V_-=0$, the two transmission identities remain valid even
when disorder causes strong elastic scattering.
For a general potential, the composite transformation relates
different disorder configurations:
\begin{equation}
\begin{split}
\mathcal T(E;\phi_1,\phi_2;V_+,V_-)
={}&\mathcal T(E;-\phi_2,-\phi_1;V_+,-V_-).
\end{split}
\label{eq:disorder_covariance}
\end{equation}
It no longer enforces a vanishing common-mode slope for each fixed
asymmetric sample. Symmetric disorder can also reduce the transmission;
stationarity alone is not a conductance guarantee.

Gate-controlled spin--orbit effects have been measured in lateral
semiconductor transport~\cite{Miller2003} and in parabolic
GaAs/AlGaAs quantum wells~\cite{Studer2009}. These experiments illustrate
the importance of electrostatic control of spin-dependent terms;
they do not supply spin--orbit parameters for the present InAs guide.

For spin-dependent extensions, the relevant sufficient condition is
covariance of the full scattering Hamiltonian under
$\mathcal A=U_{C_2}\Theta$:
\begin{equation}
\mathcal A H(\phi_1,\phi_2)\mathcal A^{-1}
\simeq H(-\phi_2,-\phi_1),
\label{eq:full_composite_condition}
\end{equation}
where $\simeq$ allows a gauge transformation and any additional
parameters are held at their specified values.
In three dimensions $C_2$ sends
$(k_x,k_y,k_z)$ to $(-k_x,-k_y,k_z)$ and
$(\sigma_x,\sigma_y,\sigma_z)$ to
$(-\sigma_x,-\sigma_y,\sigma_z)$.
Time-reversal invariance of a spin--orbit term does not by itself
establish this rotational covariance. For example, a constant
$\alpha k_z\sigma_x$ term is time-reversal even but changes sign
under $C_2$. Electric-field profiles, crystal orientation, magnetic
fields, and contacts must therefore be transformed explicitly.

Several terms relevant to InAs pass this test. The Rashba term generated
by a vertical electric field,
$\tfrac12\{\alpha(\mathbf r),k_x\sigma_y-k_y\sigma_x\}$, is
time-reversal even and invariant under $C_2$ for any vertical profile of
the field. The vertical asymmetry that produces it therefore does not by
itself break the protection; what is required is
$\alpha(C_2\mathbf r)=\alpha(\mathbf r)$, that is, equal coupling in
the two modules. More generally, a spin--orbit term of the Pauli form
$\boldsymbol\sigma\cdot(\boldsymbol\nabla V\times\mathbf p)$ derived
from a $C_2$-invariant potential satisfies
Eq.~\eqref{eq:full_composite_condition}; this includes the lateral
coupling produced by the curved walls. The bulk Dresselhaus term is
covariant when $\hat{\mathbf z}$ lies along a cubic axis such as
$[001]$, because the twofold rotation about that axis belongs to the
zinc-blende point group; for a $[110]$ or $[111]$ orientation it is not.
A difference $\alpha_-$ between the Rashba couplings of the two modules
is instead $C_2$ odd and time-reversal even. It therefore enters in the
same way as the antisymmetric potential $V_-$ of
Eq.~\eqref{eq:disorder_covariance} and restores a linear common-mode
response.

The orbital result extends to a spinful model if its complete Hamiltonian,
including any stray fields, satisfies Eq.~\eqref{eq:full_composite_condition}.

\subsection{Response to a uniform ambient field}
\label{sec:ambient}

The identities above were obtained for ideally confined fluxes, with
$\mathbf B=0$ throughout the conducting domain. That assumption excludes
the most common laboratory perturbation: a slow drift of the ambient
field in which the device is immersed. This perturbation lies in the
protected direction, and the protection does not require the drifting
field to be excluded from the conductor.

Let the ambient contribution be a spatially uniform field
$B_a\hat{\mathbf z}$ over the device, perpendicular to the plane of the
guide, and extend the parameter list to
$\mathcal T(E;B_a;\phi_1,\phi_2)$. A uniform axial field is invariant
under a proper rotation about $\hat{\mathbf z}$, so $C_2$ acts on the
extended list as in Eq.~\eqref{eq:C2},
\begin{equation}
\mathcal T(E;B_a;\phi_1,\phi_2)=\mathcal T(E;B_a;\phi_2,\phi_1),
\label{eq:ambient_C2}
\end{equation}
while microreversibility reverses every time-reversal-odd parameter,
\begin{equation}
\mathcal T(E;B_a;\phi_1,\phi_2)=\mathcal T(E;-B_a;-\phi_1,-\phi_2).
\label{eq:ambient_TRS}
\end{equation}
Their composition gives
\begin{equation}
\mathcal T(E;B_a;\phi_1,\phi_2)=\mathcal T(E;-B_a;-\phi_2,-\phi_1).
\label{eq:ambient}
\end{equation}
The drift enters the two openings with a common sign, since both signed
fluxes are referred to $+\hat{\mathbf z}$. Writing $\nu=eA_o/\hbar$ with
$A_o=\pi b_oa_o$ the opening area, the configuration at a counter-wound
bias is $(B_a;\phi+\nu B_a,-\phi+\nu B_a)$. Equation~\eqref{eq:ambient}
maps it to $(-B_a;\phi-\nu B_a,-\phi-\nu B_a)$, the same device under
the reversed drift. The counter-wound transmission is therefore even in
$B_a$ and stationary at $B_a=0$. The statement is exact within the
orbital model and holds for the field inside the conductor as well as
for the flux through the openings, because $C_2$ leaves a uniform axial
field invariant wherever it acts.

The distinction matters here because the two contributions differ
greatly in size. The aperture term is only
$\nu=6.09\,\mathrm{rad\,T^{-1}}$, whereas at $1\,\mathrm{mT}$ the
magnetic length $\sqrt{\hbar/eB_a}\simeq811\,\mathrm{nm}$ is already
comparable to $L_{\rm ref}$. In the three-dimensional calculation of
Sec.~\ref{sec:decoherence_analysis} on the $8\,\mathrm{nm}$ grid, a
uniform field of $\pm1\,\mathrm{mT}$ applied to the whole structure,
leads included, reduces the counter-wound transmission at $\phi=\pi/4$
by $2.9\times10^{-5}$ with an odd part below $10^{-12}$, whereas the
same-winding transmission changes linearly by $2.2\times10^{-3}$. A
$10\,\mu\mathrm{eV}$ potential offset in one bend restores an odd
counter-wound part of $2.4\times10^{-5}$.

Three restrictions apply. The field must be uniform on the scale of the
module separation, $|\mathbf q_1-\mathbf q_2|\simeq536\,\mathrm{nm}$: a
gradient produces unequal fluxes and hence a differential component,
which is not protected at a counter-wound bias. A Zeeman term is
compatible with Eq.~\eqref{eq:ambient}, since $\Theta$ reverses $B_a$
and $\sigma_z$ together while $C_2$ preserves both; spin--orbit terms
must instead be tested against Eq.~\eqref{eq:full_composite_condition}.
Finally, only the perpendicular component is protected. A uniform
in-plane component reverses under $C_2$ as well as under $\Theta$ and is
therefore unchanged by the composite operation: it leaves
Eq.~\eqref{eq:composite} intact at each fixed value, so the counter-wound
bias remains stationary against common offsets, but $C_2$ and time
reversal do not make the response to the in-plane component itself even.
In the present model the additional mirror symmetry $z\mapsto-z$ does
so; a substrate or a single gate removes that symmetry.

\subsection{Numerical validation}

Numerical evaluation must use the three-dimensional domain with its
elliptical through-openings and the lead boundary conditions of
Sec.~\ref{sec:model}. A discretized balanced structure must map onto
itself under $C_2$, including boundary terms and lead attachments.
Flux phases must give zero circulation on contractible loops in the
conducting region and the prescribed circulation around each opening.

A useful residual is
\begin{equation}
\epsilon_{\rm sym}
=\max_{(\phi_1,\phi_2)\in\mathcal P}
|\mathcal T(E;\phi_1,\phi_2)
-\mathcal T(E;-\phi_2,-\phi_1)|,
\label{eq:validation_residuals}
\end{equation}
where $\mathcal P$ is a set of independently evaluated flux pairs.
Gauge-equivalent choices of branch cuts and shifts of either flux
by $2\pi$ provide additional checks. A map constructed by imposing
symmetry cannot independently verify the identity used to construct it.

These algebraic checks are distinct from convergence to the
continuum domain. Lead thresholds, transmission, curvature of the
flux response, and finite drift windows must be checked under mesh
refinement at a fixed physical energy and geometry.
A small symmetry residual on a coarse mesh is not an error estimate
for the continuum conductance.

\section{Flux response and common-mode perturbations}
\label{sec:decoherence_analysis}

The symmetry relations of Sec.~\ref{sec:symmetry} constrain the phase
response but leave its magnitude undetermined. We now evaluate that
response for the three-dimensional InAs domain of Sec.~\ref{sec:model}.
The two elliptical openings extend through the full height of the guide;
each bend therefore contains two lateral arms. All results below use
$R=350\,\mathrm{nm}$, transverse semiaxes
$(a_h,a_v)=(80,60)\,\mathrm{nm}$, opening semiaxes
$(b_o,a_o)=(72.9167,17.5)\,\mathrm{nm}$, and
$E_F=7.2826087\,\mathrm{meV}$ relative to the interior conduction-band edge.
The calculation is at zero electronic temperature, with a clean scalar
potential and ideally confined fluxes.

\subsection{Numerical scattering problem}

We discretize the effective-mass Hamiltonian on a cubic grid of spacing
$a$, with nearest-neighbor kinetic coupling
$t_a=\hbar^2/(2m^*a^2)$. The outer surface and the boundaries of both
openings are evaluated from their analytic equations. A bond crossing
an excluded region is removed, including a bond whose endpoints both
lie in the conductor. If a wall intersects a missing-neighbor direction
at a distance $d<a$ from a retained site, its contribution to the onsite
kinetic term is $t_a a/d$. This distance-to-wall boundary prescription
preserves Hermiticity; its spatial error is assessed by grid refinement.

The two fluxes enter as phases on bonds crossing branch cuts from the
openings. No magnetic field is applied to the conducting volume.
Identical semi-infinite leads are attached through their retarded
self-energies. The transmission is obtained from the flux-normalized
scattering matrix, retaining all transverse degrees of freedom on the
grid, including evanescent contributions~\cite{Khomyakov2005Transport}.
Both sectors of the reflection $z\mapsto-z$ are included. Decomposing
the matrix into these sectors and reusing the zero-flux Green function
for subsequent phase values are algebraic steps; neither operation
replaces the device by a two-dimensional guide.

\begin{figure}[htbp]
\centering
\includegraphics[width=\columnwidth]{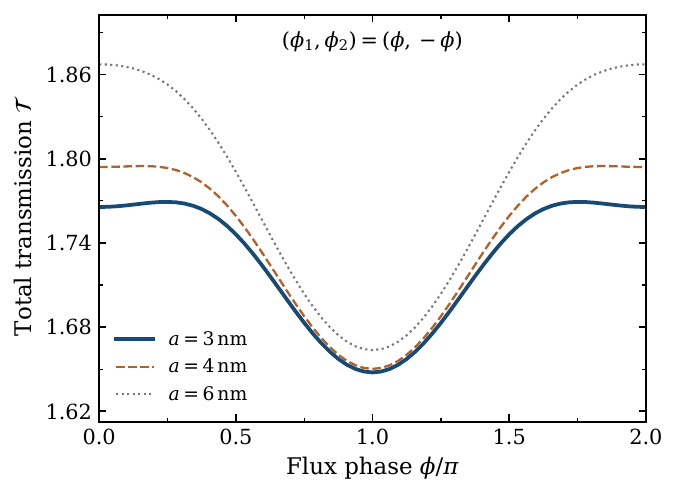}
\caption{Calculated total transmission of the balanced three-dimensional
InAs waveguide for $(\phi_1,\phi_2)=(\phi,-\phi)$ at
$E_F\simeq7.283\,\mathrm{meV}$ and zero electronic temperature.
The curves compare cubic grid spacings of $3$, $4$, and $6\,\mathrm{nm}$;
lines connect independently calculated points. The lead capacity is
three orbital channels, with spin degeneracy included only when
converting transmission to conductance. The remaining separation
between the curves is a discretization effect.}
\label{fig:CW_transmission}
\end{figure}

At the chosen energy, each lead supports three propagating orbital
modes. The first four transverse thresholds on the finest grid are
approximately $2.0731,\,4.5169,\,5.9852,\,8.0533\,\mathrm{meV}$.
For a straight guide with the same cross section, the $10\,\mathrm{nm}$
grid reproduces $\mathcal T=3$ to within $10^{-12}$. Flux periodicity, reciprocity, exchange of the
modules, and current conservation are checked separately from spatial
convergence. Over the 129 common-offset pairs of
Sec.~\ref{sec:static_offset} ($a=3\,\mathrm{nm}$, $\phi=\pi/4$), the
symmetry residual $\epsilon_{\rm sym}$ defined in Sec.~\ref{sec:symmetry}
is below $4\times10^{-14}$.
These algebraic checks do not establish continuum accuracy.
The discretization described here was also implemented independently
from this description alone. That implementation reproduces the
counter-wound sweeps of Fig.~\ref{fig:CW_transmission} at $a=6$ and
$4\,\mathrm{nm}$ at all 65 phases to within $10^{-6}$, the
$a=8\,\mathrm{nm}$ entries of Table~\ref{tab:response_mesh}, the lead
thresholds quoted above, and the flux-mismatch coefficients
$b_\varepsilon$ of Table~\ref{tab:robustness_mesh}. It is used for the
uniform-field check of Sec.~\ref{sec:ambient} and for the arm-spectrum,
energy-resolved, grid-alignment, and finite-temperature results below.

\subsection{Counter-wound phase sweep}
\label{sec:cw_sweep}

Figure~\ref{fig:CW_transmission} shows
$\mathcal T_{\rm CW}(\phi)=\mathcal T(E_F;\phi,-\phi)$ over one flux period.
Every plotted phase point is evaluated independently; the curve is not
constructed by reflecting half a period or by fitting a sinusoid.
The response is $2\pi$-periodic and satisfies
$\mathcal T_{\rm CW}(2\pi-\phi)=\mathcal T_{\rm CW}(\phi)$.
It remains flux dependent despite $\Phi_1+\Phi_2=0$, in agreement with
the winding argument of Sec.~\ref{sec:phases}.

On the $a=3\,\mathrm{nm}$ grid, the sampled transmission ranges from
$1.648$ to $1.769$. The largest sampled values lie near $\phi/\pi\simeq0.25$ and $1.75$, while the smallest occurs at $\phi=\pi$. The weak displacement of the maxima from zero phase is sensitive to grid
refinement: the maximum lies at $\phi=0$ on the $8$ and $6\,\mathrm{nm}$
grids and moves to $\phi/\pi\simeq0.16$ and $0.25$ at $4$ and
$3\,\mathrm{nm}$. Both extrema are therefore taken from the sampled sweep
rather than from $\mathcal T(0)$ and $\mathcal T(\pi)$. We quote two
conventional measures of the modulation, the visibility and the relative
modulation depth:
\begin{align}
\mathcal V&=\frac{\mathcal T_{\max}-\mathcal T_{\min}}
{\mathcal T_{\max}+\mathcal T_{\min}},\label{eq:response_visibility}\\
\mathcal D&=\frac{\mathcal T_{\max}-\mathcal T_{\min}}
{\mathcal T_{\max}}.\label{eq:response_depth}
\end{align}
The corresponding finite-grid values are
$\mathcal V\simeq3.56\%$ and $\mathcal D\simeq6.87\%$. They describe the
same pair of extrema and are related by $\mathcal V=\mathcal D/(2-\mathcal D)$.
Neither percentage is the fraction of the three-channel lead capacity.
Using $G=(2e^2/h)\mathcal T$, the same sweep gives
$G\simeq127.7\text{--}137.1\,\mu\mathrm S$ and a two-terminal resistance
of approximately $7.29\text{--}7.83\,\mathrm{k}\Omega$.
This resistance includes the quantum contact contribution.

The three open lead modes do not contribute equally to transmission.
At the midpoint of an opening each arm is an elliptical segment
$a_o<u<a_h$ of the transverse cross section. Its lowest level,
$6.723\,\mathrm{meV}$ in the continuum, lies only $0.56\,\mathrm{meV}$
below $E_F$, and its first level odd under $z\mapsto-z$ lies
$5.6\,\mathrm{meV}$ above $E_F$. Each arm therefore carries a single open
mode. Because the $z$ parity is conserved, the transmission separates
as $\mathcal T=\mathcal T^{(+)}+\mathcal T^{(-)}$, the sum of the sectors
even and odd under $z\mapsto-z$. The even sector contains two lead modes,
so $\mathcal T^{(+)}\le2$. The odd third lead mode is evanescent across
each opening, with a tunneling factor of order $10^{-8}$; on the
$8\,\mathrm{nm}$ grid the calculated $\mathcal T^{(-)}$ is below
$10^{-12}$. Apart from this negligible
odd contribution, $\mathcal T\le2$. The loss of transmitted
current is accounted for by reflection; no loss of coherence is
introduced in this calculation.

\begin{figure*}[htbp]
\centering
\includegraphics[width=\textwidth]{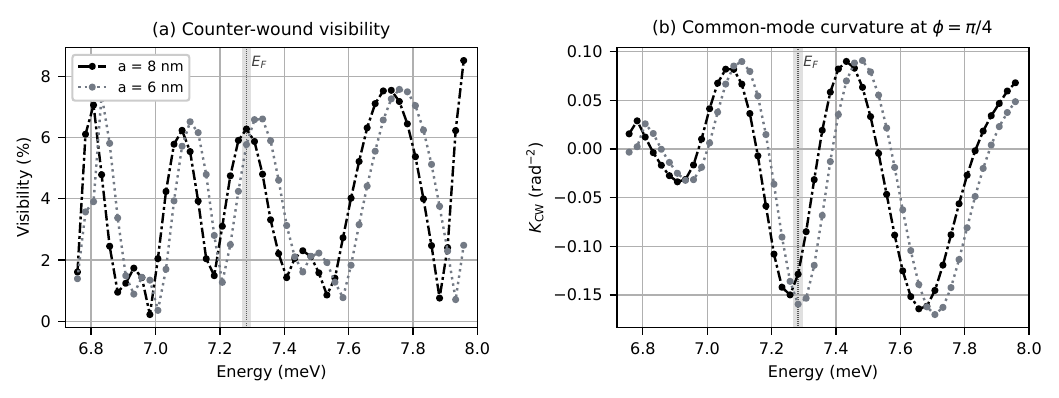}
\caption{Energy dependence of the counter-wound response on the $8$ and
$6\,\mathrm{nm}$ grids at zero electronic temperature. (a) Visibility of
the counter-wound sweep $\mathcal T(E;\phi,-\phi)$ over one flux period.
(b) Common-mode curvature $K_{\rm CW}$ at $\phi=\pi/4$. The dotted
vertical line marks $E_F$; the shaded band is the full width at half
maximum of $-\partial f/\partial E$ at $100\,\mathrm{mK}$. The two grids
differ mainly by a rigid shift of about $35\,\mu\mathrm{eV}$
(Sec.~\ref{sec:cw_sweep}).}
\label{fig:energy}
\end{figure*}

Reducing the grid spacing from $4$ to $3\,\mathrm{nm}$ changes the
sampled transmission by at most $0.028$, or $1.61\%$ relative to the
finer-grid value at the same phase. The largest change from $6$ to
$4\,\mathrm{nm}$ is $0.073$, and the visibility changes by approximately
$18\%$ between the two finest grids. Most of this grid dependence is a
displacement in energy rather than a change in the shape of the
response. The discretization lowers the arm threshold by approximately
$95$, $53$, $24$, and $13\,\mu\mathrm{eV}$ at $a=8$, $6$, $4$, and
$3\,\mathrm{nm}$, and the energy structure of the transmission follows
it. Over $6.76$--$7.96\,\mathrm{meV}$ a single rigid shift of
$35\,\mu\mathrm{eV}$ maps $\mathcal T(E;\phi,-\phi)$ on the
$8\,\mathrm{nm}$ grid onto the $6\,\mathrm{nm}$ result, reducing the
root-mean-square difference over all phases and energies from $0.097$
to $0.014$; the corresponding threshold displacement is
$42\,\mu\mathrm{eV}$. Because the visibility varies steeply with energy
near $E_F$ [Fig.~\ref{fig:energy}(a)], comparing grids at a fixed $E_F$
mixes this shift with genuine convergence.

Comparing the grids instead at energies displaced by $0.83$ of each
grid's arm-threshold displacement---the ratio fixed by the $8$ and
$6\,\mathrm{nm}$ spectra---gives $\mathcal V=2.83$, $2.92$, and $2.81\%$
at $a=8$, $6$, and $4\,\mathrm{nm}$; the $4\,\mathrm{nm}$ value was not
used to fix the ratio and therefore tests it. Referring each grid to its
full threshold displacement gives $1.78$, $2.31$, and $2.54\%$.
On the $3\,\mathrm{nm}$ grid the two references give $2.80$ and $2.65\%$,
and their difference decreases approximately as $a^2$, from $1.05$,
$0.61$, and $0.27$ to $0.15$ percentage points. The two references must
coincide as $a\to0$; we take the range they span, $2.6$--$2.9\%$, as the
estimate of the continuum visibility at the nominal $E_F$, below the
finest-grid value of $3.56\%$. The same comparison gives
$K_{\rm CW}=-0.093$ to $-0.101\,\mathrm{rad}^{-2}$,
$S_{\rm SW}=-0.051$ to $-0.057\,\mathrm{rad}^{-1}$, and
$b_\varepsilon=0.021$ to $0.023\,\mathrm{rad}^{-1}$ at $\phi=\pi/4$, to be
compared with the $3\,\mathrm{nm}$ values of
Table~\ref{tab:response_derivatives} and Eq.~\eqref{eq:imbalance_slope}.
The finite-grid numbers quoted in this section and in
Sec.~\ref{sec:robustness} should be read with this systematic offset in
mind; a direct confirmation requires an energy-resolved calculation on a
finer grid.

A second, smaller source of grid dependence is the alignment of the
grid with the curved walls. Shifting the grid origin by half a cell
along $x$ and/or $y$, which preserves the $C_2$ symmetry exactly, changes
the visibility by up to $0.63$, $0.16$, and $0.08$ percentage points at
$a=8$, $6$, and $4\,\mathrm{nm}$. On the two finer grids this is small
compared with the energy displacement described above.

\begin{table}[htbp]
\centering
\caption{Grid dependence of the counter-wound sweep at fixed physical
geometry and Fermi energy. $\mathcal T_{\max}$ and $\mathcal T_{\min}$
are the extrema of the sampled sweep, from which the visibility and depth
follow via Eqs.~\eqref{eq:response_visibility} and
\eqref{eq:response_depth}. The minimum lies at $\phi=\pi$ on every grid;
the maximum lies at $\phi=0$ on the $8$ and $6\,\mathrm{nm}$ grids only.
These are finite-grid results at fixed $E_F$, not a continuum
extrapolation (see Sec.~\ref{sec:cw_sweep}).}
\label{tab:response_mesh}
\begin{tabular}{cccccc}
\toprule
$a$ (nm) & $\mathcal T(0)$ & $\mathcal T_{\max}$ & $\mathcal T_{\min}$ & $\mathcal V$ (\%) & $\mathcal D$ (\%)\\
\midrule
8 & 1.9404 & 1.9404 & 1.7112 & 6.28 & 11.81 \\
6 & 1.8674 & 1.8674 & 1.6636 & 5.77 & 10.91 \\
4 & 1.7941 & 1.7948 & 1.6502 & 4.20 & 8.06 \\
3 & 1.7656 & 1.7693 & 1.6477 & 3.56 & 6.87 \\
\bottomrule
\end{tabular}
\end{table}

\begin{table}[htbp]
\centering
\caption{Local common-offset response on the $a=3\,\mathrm{nm}$ grid.
The slope is in $\mathrm{rad}^{-1}$ and the curvature in
$\mathrm{rad}^{-2}$. Central differences use $\Delta c=0.005\,\mathrm{rad}$;
steps of $0.01$ and $0.02\,\mathrm{rad}$ check the differentiation.
At $\phi=\pi/4$ the same-winding curvature is
$K_{\rm SW}=-0.0381\,\mathrm{rad}^{-2}$.
Spatial discretization remains a separate source of uncertainty.}
\label{tab:response_derivatives}
\begin{tabular}{ccccc}
\toprule
$\phi$ & $T_{{\rm CW}0}$ & $T_{{\rm SW}0}$ & $K_{\rm CW}$ & $S_{\rm SW}$\\
\midrule
$\pi/4$ & 1.769 & 1.733 & -0.120 & -0.071 \\
$\pi/2$ & 1.746 & 1.678 & -0.088 & -0.057 \\
$3\pi/4$ & 1.683 & 1.651 & -0.025 & -0.014 \\
\bottomrule
\end{tabular}
\end{table}

\subsection{Energy dependence}
\label{sec:energy}

The symmetry identities hold at every energy, but the magnitudes they
leave free do not. Figure~\ref{fig:energy} shows the counter-wound
response between the arm threshold and the opening of the fourth lead
mode, calculated on the $8$ and $6\,\mathrm{nm}$ grids. Between $6.76$
and $7.96\,\mathrm{meV}$ the visibility varies between $0.4$ and $7.6\%$
on the $6\,\mathrm{nm}$ grid, with neighboring maxima
$0.13$--$0.25\,\mathrm{meV}$ apart, and the transmission at $\phi=0$
between $1.09$ and $1.97$. The common-mode curvature $K_{\rm CW}$ at
$\phi=\pi/4$ changes sign repeatedly; on the $6\,\mathrm{nm}$ grid the
nearest sign changes lie $93\,\mu\mathrm{eV}$ below and
$107\,\mu\mathrm{eV}$ above $E_F$. At those energies the counter-wound
bias is stationary to fourth order in the common offset, while between
them the common-offset response alternates between a maximum and a
minimum. The drift tolerances of Sec.~\ref{sec:static_offset} are
therefore properties of the selected energy rather than of the geometry
alone. At $E_F$ the local counter-wound half-width of
Eq.~\eqref{eq:response_local_windows} exceeds the same-winding one by a
factor of $2.9$ on the $6\,\mathrm{nm}$ grid, but $75\,\mu\mathrm{eV}$
lower the same-winding assignment is the more tolerant of the two. These
energy scales are only a few times the $30\,\mu\mathrm{eV}$ thermal width
at $100\,\mathrm{mK}$.

\subsection{A common static phase offset}
\label{sec:static_offset}

At a prescribed bias phase $\phi$, define
\begin{align}
\mathcal T_{\rm CW}(c)&=\mathcal T(E_F;\phi+c,-\phi+c),\\
\mathcal T_{\rm SW}(c)&=\mathcal T(E_F;\phi+c,\phi+c).
\end{align}
For each configuration, let $T_{\alpha0}=\mathcal T_\alpha(0)$.
The two baseline transmissions are generally different. With
$K_\alpha=\partial_c^2\mathcal T_\alpha|_0$ and
$S_{\rm SW}=\partial_c\mathcal T_{\rm SW}|_0$, the local responses are
\begin{align}
\mathcal T_{\rm CW}(c)&=T_{{\rm CW}0}+\tfrac12K_{\rm CW}c^2+O(c^4),\\
\mathcal T_{\rm SW}(c)&=T_{{\rm SW}0}+S_{\rm SW}c
+\tfrac12K_{\rm SW}c^2+O(c^3).
\end{align}
The absent linear term in the first expression follows from
Sec.~\ref{sec:symmetry}. It does not determine the sign or magnitude
of $K_{\rm CW}$.

For a fractional tolerance $\epsilon$, the leading local estimates are
\begin{equation}
c_{\rm CW}^{\rm loc}=\sqrt{\frac{2\epsilon T_{{\rm CW}0}}{|K_{\rm CW}|}},
\qquad
c_{\rm SW}^{\rm loc}=\frac{\epsilon T_{{\rm SW}0}}{|S_{\rm SW}|}.
\label{eq:response_local_windows}
\end{equation}
They apply only while the retained Taylor terms describe the response.
When a denominator is small, higher derivatives must be included and
the actual tolerance must be found from the full transmission curve.
The symmetry alone supplies no universal numerical enhancement factor.

We test the local estimates with a full common-offset sweep at
$\phi=\pi/4$ on the $a=3\,\mathrm{nm}$ grid, using the same
three-dimensional scattering model as the preceding calculations.
Each of the 129 offsets in $-\pi\le c\le\pi$ is evaluated
independently for both configurations. The differential coordinate
remains fixed: $D=\phi$ for CW and $D=0$ for SW, while
$C=c$ and $C=\phi+c$, respectively. No reflection of the calculated
curve is imposed. The maximum sampled difference
$|\mathcal T_{\rm CW}(c)-\mathcal T_{\rm CW}(-c)|$
is below $4\times10^{-14}$.

For the finite-offset comparison, the tolerance is defined by
\[
\frac{|\mathcal T_\alpha(c)-T_{\alpha0}|}{T_{\alpha0}}
\le \epsilon
\]
throughout the connected interval containing $c=0$.
An increase and a decrease of equal magnitude therefore count
as the same signal deviation.

At $\epsilon=0.01$, additional scattering evaluations locate the
first threshold crossings at approximately
$c=\pm0.5679\,\mathrm{rad}$ for CW. For SW, the crossings occur
at $c=-0.2711\,\mathrm{rad}$ and $c=+0.2341\,\mathrm{rad}$.
The largest allowed symmetric half-widths are consequently
$0.5679$ and $0.2341\,\mathrm{rad}$.
Equation~\eqref{eq:response_local_windows} predicts
$0.5438$ and $0.2438\,\mathrm{rad}$, respectively, differing
from the full-response half-widths by approximately $4.2\%$.

The ratio of full half-widths is approximately $2.43$ for this
tolerance and bias. It is not a universal enhancement factor:
at $\epsilon=0.001$, the corresponding half-widths are
$0.1727$ and $0.02423\,\mathrm{rad}$.
These finite-offset comparisons retain the spatial discretization
uncertainty and zero-temperature assumptions of the calculation.
They do not establish continuum drift tolerances or their values
at other energies.

\begin{figure}[b]
    \centering
    \includegraphics[
        width=\columnwidth
    ]{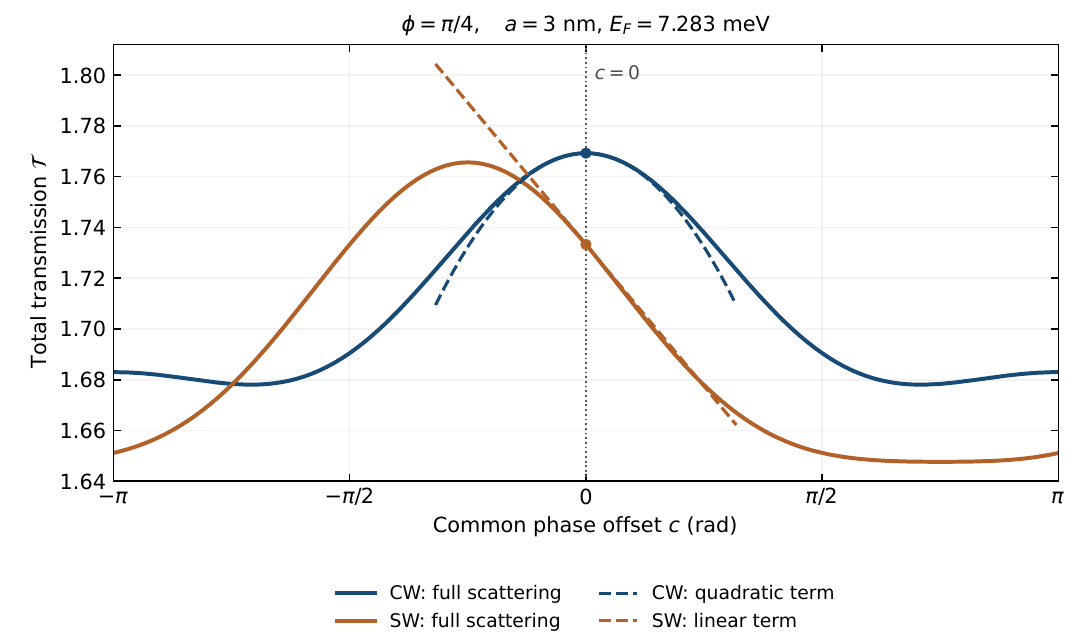}
    \caption{
        Common-offset response at $\phi=\pi/4$ for the
        counter-wound (CW) and same-winding (SW)
        configurations on the $3\,\mathrm{nm}$ grid.
        Solid curves connect independently calculated
        transmission values. Dashed curves show the
        leading local approximations: quadratic for CW
        and linear for SW, displayed for
        $|c|\leq 1\,\mathrm{rad}$.
        The vertical line marks $c=0$.
        Calculations use $E_F=7.2826087\,\mathrm{meV}$
        and zero electronic temperature.
    }
    \label{fig:common_offset_sweep}
\end{figure}

\begin{figure*}[htbp]
\centering
\includegraphics[width=\textwidth]{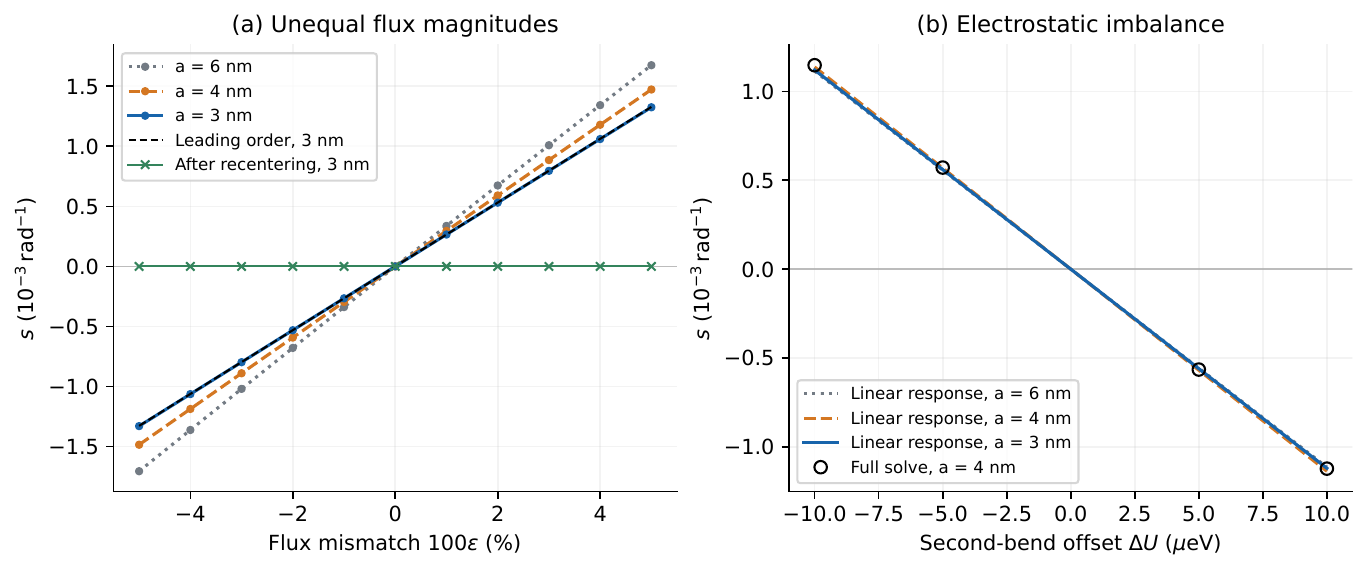}
\caption{Common-mode sensitivity at $\phi=\pi/4$.
(a) Normalized slope at the nominal operating point for unequal flux
magnitudes. Markers are full scattering calculations; lines join them.
The thin black dashed line is the leading expression
\eqref{eq:imbalance_slope} on the $3\,\mathrm{nm}$ grid; green crosses
show the recalculated slopes after recentering.
(b) Response to a uniform potential offset in the second bend, with
both leads and the first bend unchanged. Lines are first-order
potential-response predictions for three grid spacings. Open markers
are separate finite-offset scattering calculations on the
$4\,\mathrm{nm}$ grid. Their agreement tests the local expansion, not
convergence to the continuum. Both panels use the transmission of the
perturbed configuration to define the normalized slope.}
\label{fig:robustness_asymmetry}
\end{figure*}

The full responses and their leading local approximations are compared
in Fig.~\ref{fig:common_offset_sweep}.

\subsection{A fluctuating common offset}

For quasistatic, zero-mean Gaussian fluctuations, take the same offset
$c$ at both openings, with $\langle c^2\rangle=\sigma_c^2$.
This is a purely common perturbation. It differs from independent
fluctuations of the two fluxes, which also contain a differential component.
The small-noise moments follow from Eq.~\eqref{eq:CW_noise_moments}.
For example, at $\phi=\pi/4$ the finest-grid derivatives give
\begin{align}
\frac{\langle\mathcal T_{\rm CW}\rangle}{T_{{\rm CW}0}}
&=1-0.0338\,\sigma_c^2+O(\sigma_c^4),\\
\frac{\langle\mathcal T_{\rm SW}\rangle}{T_{{\rm SW}0}}
&=1-0.0110\,\sigma_c^2+O(\sigma_c^4),\\
\frac{\operatorname{Var}(\mathcal T_{\rm CW})}{T_{{\rm CW}0}^2}
&=0.00229\,\sigma_c^4+O(\sigma_c^6),\\
\frac{\operatorname{Var}(\mathcal T_{\rm SW})}{T_{{\rm SW}0}^2}
&=0.00168\,\sigma_c^2+O(\sigma_c^4),
\end{align}
where $\sigma_c$ is expressed in radians. These coefficients are derived
from the local scattering response; they are not Monte Carlo estimates
at a finite noise amplitude. They share the spatial uncertainty of the
underlying derivatives.

The symmetry suppresses the leading transmission fluctuations relative
to a same-winding bias with nonzero slope. The mean shift remains
quadratic in the noise amplitude in both configurations, and its ordering
depends on their curvatures. These statements concern an ensemble of
static coherent scattering problems. They do not include dynamical
dephasing or thermal energy averaging.

Environment-induced loss of interference and models of dephasing in
Aharonov--Bohm rings address additional physics~\cite{Stern1990,Benjamin2002}.
In the present quasistatic ensemble, no absorption or reinjection mechanism is introduced.

\section{Asymmetry and disorder}
\label{sec:robustness}

The common-mode stationarity derived in Sec.~\ref{sec:symmetry}
requires a balanced scattering structure. Here we distinguish a mismatch
between the two fluxes from an asymmetry of the scalar potential. The
former displaces the operating point in flux space; the latter changes
the Hamiltonian itself. We use the three-dimensional domain, Fermi
energy, and zero-temperature scattering calculation of
Sec.~\ref{sec:decoherence_analysis}, with a bias phase $\phi=\pi/4$.

For any fixed sample, define the normalized common-mode slope
\begin{equation}
 s=\left.\frac{1}{\mathcal T(0)}
 \frac{\partial\mathcal T(c)}{\partial c}\right|_{c=0}.
 \label{eq:robustness_slope}
\end{equation}
The denominator is the transmission of that same sample and flux
assignment. Thus $s$ is a fractional sensitivity, in
$\mathrm{rad}^{-1}$, rather than a transmission probability.
In the clean balanced device, $s_{\rm CW}=0$, while the finest grid gives
$s_{\rm SW}=-0.0410\,\mathrm{rad}^{-1}$.
These slopes describe response to a subsequent common flux perturbation;
they do not measure the transmission lost to an imperfection.

\begin{figure*}[htbp]
\centering
\includegraphics[width=\textwidth]{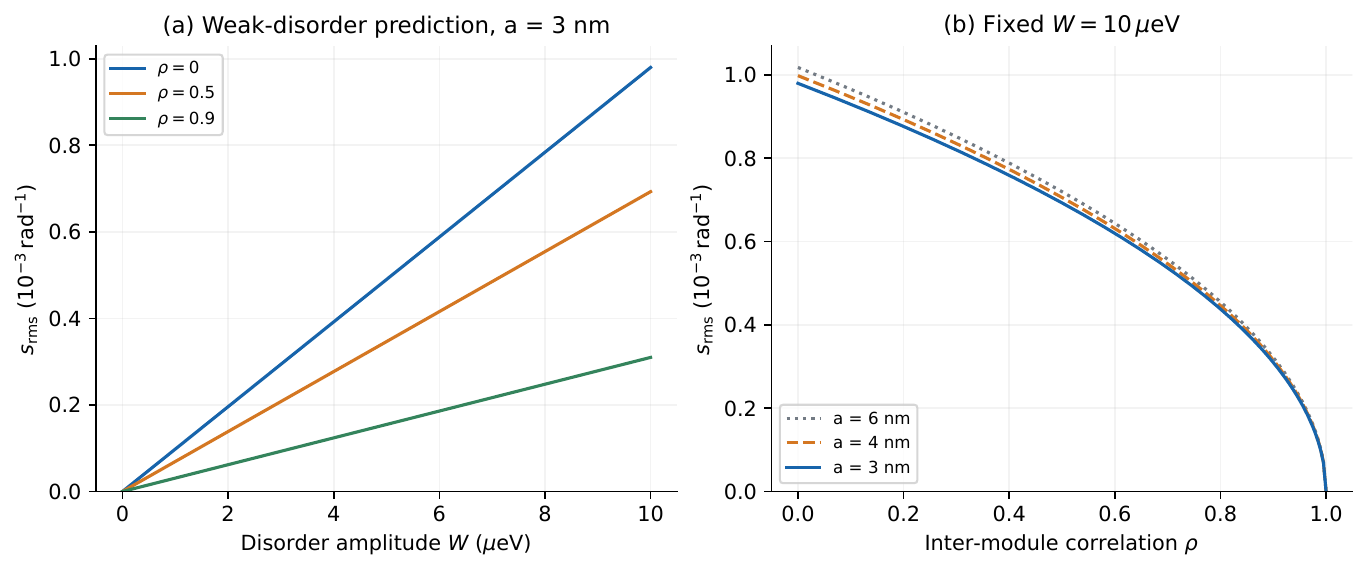}
\caption{Weak-disorder predictions for the three-dimensional scalar
potential model with $\xi=20\,\mathrm{nm}$.
(a) Rms common-mode slope versus disorder amplitude for three
correlations between the modules, using the $3\,\mathrm{nm}$ response
kernel. (b) Correlation dependence at $W=10\,\mu\mathrm{eV}$, comparing
three grid spacings at fixed physical covariance.
All curves evaluate Eq.~\eqref{eq:weak_disorder_rms}; they are not
finite-disorder ensemble simulations. The mesh comparison exposes the
remaining numerical uncertainty in the coefficient $A_\xi$.}
\label{fig:robustness_disorder}
\end{figure*}

\subsection{Unequal flux magnitudes}

Let the second flux differ in magnitude by a fraction $\varepsilon$:
\begin{equation}
 \phi_1=\phi+c,\qquad
 \phi_2=-(1+\varepsilon)\phi+c.
 \label{eq:flux_imbalance}
\end{equation}
The coordinates of Sec.~\ref{sec:symmetry} become
\begin{equation}
 C=c-\frac{\varepsilon\phi}{2},\qquad
 D=\phi\left(1+\frac{\varepsilon}{2}\right).
 \label{eq:imbalance_coordinates}
\end{equation}
At the nominal setting $c=0$, the device therefore lies away from the
stationary line $C=0$. Expanding the normalized slope about the balanced
point gives
\begin{equation}
 s_\varepsilon=-\frac{\phi K_c}{2T_0}\varepsilon
 +O(\varepsilon^2),
 \label{eq:imbalance_slope}
\end{equation}
where $T_0=F_{E_F}(0,\phi)$ and
$K_c=\partial_C^2F_{E_F}(0,\phi)$ refer to the clean CW device.
The $3\,\mathrm{nm}$ grid gives
$-\phi K_c/(2T_0)=0.0266\,\mathrm{rad}^{-1}$.
A $+5\%$ mismatch gives
$s_\varepsilon=1.325\times10^{-3}\,\mathrm{rad}^{-1}$
from the full phase-dependent scattering calculation.

Provided the scalar structure remains symmetric, setting
\begin{equation}
 c_*=\frac{\varepsilon\phi}{2}
 \label{eq:flux_recentring}
\end{equation}
restores common-mode stationarity exactly. At a $5\%$ mismatch this
requires $c_*\simeq0.0196\,\mathrm{rad}$. The differential coordinate
has also changed, so restoring zero slope does not in general restore
the original transmission. The calculated residual slopes after
recentering are below $10^{-10}\,\mathrm{rad}^{-1}$ on the grids shown.
No mechanical tilt tolerance is inferred: $\varepsilon$ parametrizes
the linked fluxes, not the orientation of a finite magnetic source.

\subsection{Electrostatic imbalance between the bends}

We next keep $(\phi_1,\phi_2)=(\phi+c,-\phi+c)$ and add a uniform
onsite potential $\Delta U$ inside the second bend. The first bend and
both semi-infinite leads remain at zero potential. This includes the
scattering caused by the potential steps at the bend boundaries.
Writing the resulting slope as
\begin{equation}
 s(\Delta U)=\chi_U\Delta U+O(\Delta U^2),
 \label{eq:potential_imbalance}
\end{equation}
we find $\chi_U=-0.1121\,\mathrm{meV}^{-1}\mathrm{rad}^{-1}$
on the finest grid. Thus a $+10\,\mu\mathrm{eV}$ offset has the
first-order estimate $s=-1.121\times10^{-3}\,\mathrm{rad}^{-1}$.
The finite-offset checks in Fig.~\ref{fig:robustness_asymmetry}(b)
agree with the corresponding $4\,\mathrm{nm}$ linear prediction to
within 1.31\% over the four tested nonzero offsets,
$\Delta U=\pm5,\pm10\,\mu\mathrm{eV}$.

A scalar potential offset is not equivalent to a geometric change of
arm length. It changes the local longitudinal wave numbers, junction
scattering, and mode conversion. We therefore express this result in
energy units and do not convert it into a nanometer fabrication tolerance.
A change of the opening dimensions or bend radius would require a
separate domain calculation.

\subsection{Spatially correlated scalar disorder}

To make the disorder model independent of grid spacing, we prescribe
its correlation length in physical units. On the first bend, let
$g_1(\mathbf r)$ and $g_2(\mathbf r)$ be independent, zero-mean Gaussian
fields with unit variance and covariance
\begin{equation}
 \langle g_\alpha(\mathbf r)g_\beta(\mathbf r')\rangle
 =\delta_{\alpha\beta}
 \exp\left[-\frac{|\mathbf r-\mathbf r'|^2}{2\xi^2}\right].
 \label{eq:disorder_spatial_covariance}
\end{equation}
We choose $\xi=20\,\mathrm{nm}$ as a specified model parameter, not
as a measured material property. For a point $\mathbf r$ in the first
bend, the potential in the two modules is defined by
\begin{align}
 V(\mathbf r)&=Ww(\mathbf r)g_1(\mathbf r),\notag\\
 V(C_2\mathbf r)&=Ww(\mathbf r)
 \left[\rho g_1(\mathbf r)+\sqrt{1-\rho^2}\,g_2(\mathbf r)\right],
 \label{eq:paired_disorder}
\end{align}
Here $W$ is an energy scale, and $0\leq\rho\leq1$ is the correlation
between potentials at symmetry-related points. The envelope
$w(\mathbf r)=\sin^2(2\theta)$, with
$\theta=\operatorname{atan2}(-x,R-y)$ in the first bend, vanishes at
the central junction and at the lead interface. It is reflected into
the second bend, and the lead potentials remain zero. The local rms
potential is $Ww(\mathbf r)$ in either module. The fields vary in all
three spatial coordinates; no transverse averaging is imposed on them.

For weak disorder, expand the slope in the onsite potentials of the
three-dimensional grid:
\begin{equation}
 s[V]=\sum_i k_iV_i+O(V^2),\qquad
 k_i=\left.\frac{\partial s}{\partial V_i}\right|_{V=0}.
 \label{eq:disorder_kernel}
\end{equation}
The coefficients are obtained by differentiating the retarded
Green-function scattering expression~\cite{Khomyakov2005Transport}.
In particular, with
$q_i(c)=\partial\mathcal T(c)/\partial V_i|_{V=0}$,
\begin{equation}
 k_i=\frac{1}{T_0}\left.\frac{\partial q_i(c)}{\partial c}\right|_0
 \quad\text{for the balanced CW configuration}.
 \label{eq:kernel_derivative}
\end{equation}
Both reflection sectors of the clean three-dimensional Hamiltonian are
retained when evaluating this derivative. The covariance relation of
Sec.~\ref{sec:symmetry} implies
$k_{C_2i}=-k_i$: only the antisymmetric part of a weak scalar
perturbation contributes to the first-order slope.

Let $i,j$ now run over the first bend, and define
\begin{equation}
 A_\xi^2=2\sum_{i,j}k_iw_i k_jw_j
 \exp\left[-\frac{|\mathbf r_i-\mathbf r_j|^2}{2\xi^2}\right].
 \label{eq:disorder_coefficient}
\end{equation}
Averaging the linear response over the Gaussian fields gives
\begin{equation}
 s_{\rm rms}\equiv\sqrt{\langle s^2\rangle}
 =A_\xi W\sqrt{1-\rho}+o(W).
 \label{eq:weak_disorder_rms}
\end{equation}
For $a=3\,\mathrm{nm}$ and $\xi=20\,\mathrm{nm}$,
$A_\xi=0.0980\,\mathrm{meV}^{-1}\mathrm{rad}^{-1}$.
At $W=10\,\mu\mathrm{eV}$ and $\rho=0$, the corresponding prediction
is $s_{\rm rms}=9.798\times10^{-4}\,\mathrm{rad}^{-1}$.
The square-root dependence in Eq.~\eqref{eq:weak_disorder_rms}
follows from the specified covariance and the calculated antisymmetric
kernel. It is not an exponent extracted from finite-disorder ensembles.
For $\rho=1$, each potential realization is exactly $C_2$ symmetric:
$s=0$ then follows from symmetry at any disorder strength for which
the normalized slope is defined, although the transmission can decrease.

\subsection{Validation and limits of the estimates}

The common-phase derivative is evaluated with a central step of
$0.005\,\mathrm{rad}$ and checked against $0.01\,\mathrm{rad}$.
On the finest grid the relative change in the CW kernel is below
0.002\%, and its antisymmetry residual is below
$10^{-8}$ in relative Euclidean norm. These checks concern the
response calculation on a fixed grid.

As a separate check, the full scattering problem was solved for one
three-dimensional disorder pattern at $W=5$ and
$10\,\mu\mathrm{eV}$ with $\rho=0$ on the $6\,\mathrm{nm}$ grid,
without using reflection-sector reduction. The predicted and directly
calculated slopes differ by 0.36\% and 0.71\%,
respectively. A symmetric pattern with $\rho=1$ at
$W=10\,\mu\mathrm{eV}$ gives $|s|<10^{-10}\,\mathrm{rad}^{-1}$.
These selected checks support the local response calculation but do
not establish an ensemble-wide bound on nonlinear corrections.

\begin{table}[t]
\centering
\caption{Spatial convergence of the normalized sensitivity coefficients
at $\phi=\pi/4$. The flux coefficient $b_\varepsilon$ is defined by
$s_\varepsilon=b_\varepsilon\varepsilon+O(\varepsilon^2)$.
The disorder coefficient uses $\xi=20\,\mathrm{nm}$.}
\label{tab:robustness_mesh}
\begin{tabular}{cccc}
\toprule
$a\;(\mathrm{nm})$ & $b_\varepsilon$ & $\chi_U$ & $A_\xi$\\
 & $\mathrm{rad}^{-1}$
 & $\mathrm{meV}^{-1}\mathrm{rad}^{-1}$
 & $\mathrm{meV}^{-1}\mathrm{rad}^{-1}$\\
\midrule
6 & 0.03382 & -0.11149 & 0.10179 \\
4 & 0.02959 & -0.11379 & 0.09982 \\
3 & 0.02656 & -0.11212 & 0.09798 \\
\bottomrule
\end{tabular}
\end{table}

Table~\ref{tab:robustness_mesh} also shows that spatial uncertainty is
larger than the differentiation error. Between the $4$ and
$3\,\mathrm{nm}$ grids, $b_\varepsilon$, $\chi_U$, and $A_\xi$
change by approximately $11.4\%$, $1.49\%$, and $1.88\%$, respectively,
relative to the finest-grid magnitudes.
The quoted sensitivities should therefore be read as finite-grid
estimates; for $b_\varepsilon$ the grid dependence has the origin
discussed in Sec.~\ref{sec:cw_sweep}. The stationary line and the distinction between symmetric
and antisymmetric perturbations are exact within the stated orbital
model; their numerical coefficients remain geometry and energy dependent.

The disorder average above describes variations between static coherent
samples. If a given imperfect sample is subsequently exposed to a small
common phase fluctuation of variance $\sigma_c^2$, its nonzero slope
contributes $s^2\sigma_c^2$ to the leading fractional transmission
variance. The absence of this term in a perfectly symmetric sample is
the protection studied here; it is not a statement about a dephasing rate.

No mean free path is assigned to $W$ here. Such a conversion would
require a separate transport calibration using the same material,
covariance, and confinement. Consequently these calculations do not
supply a mobility threshold or a universal fabrication tolerance.
Inelastic scattering, spin-dependent terms, and magnetic leakage are
also outside the present calculation; thermal averaging of the clean
response is treated in Sec.~\ref{sec:temperature}.

\section{Experimental feasibility}
\label{sec:feasibility}

The calculations above specify an orbital transport model, rather than a
completed fabrication design. An experimental realization must combine
few-mode conduction, phase coherence, matched electrostatic confinement,
and two independently controlled flux sources. We assess these requirements
using the geometry of Sec.~\ref{sec:model}; numerical transmission values
retain the finite-grid qualifications of
Secs.~\ref{sec:decoherence_analysis} and \ref{sec:robustness}, and finite
temperature is treated in Sec.~\ref{sec:temperature}.

\subsection{Confinement and carrier occupation}

The transverse semiaxes are $a_h=80\,\mathrm{nm}$ and
$a_v=60\,\mathrm{nm}$, giving a full cross section of
$160\times120\,\mathrm{nm}$. The bend radius is $R=350\,\mathrm{nm}$,
and the reference length is $L_{\rm ref}=\pi R=1.100\,\mu\mathrm m$,
excluding the leads. Each opening has full in-plane dimensions
$145.83\times35\,\mathrm{nm}$ and passes through the entire guide height.
The opening length is a design parameter of the reference geometry,
not a measured fabrication dimension.

For $m^*=0.023m_e$ and $E_F=7.2826\,\mathrm{meV}$, the bulk-equivalent
wave number $k_0=\sqrt{2m^*E_F}/\hbar$ gives
$\lambda_0=2\pi/k_0=94.76\,\mathrm{nm}$ and
$v_0=\hbar k_0/m^*=3.34\times10^5\,\mathrm{m\,s^{-1}}$.
These are useful energy scales, but propagation in the leads is governed
by the occupied subbands. A separate variational solution of the
continuum hard-wall transverse ellipse gives the first four thresholds
\begin{equation}
 (E_1^\perp,E_2^\perp,E_3^\perp,E_4^\perp)
 \simeq(2.074,4.523,5.998,8.075)\,\mathrm{meV}.
 \label{eq:feasibility_thresholds}
\end{equation}
These continuum lead estimates differ slightly from the discretized
thresholds used in the full scattering calculation. Both descriptions
have three open orbital modes at the selected energy. For each mode,
\begin{equation}
 k_n=\frac{\sqrt{2m^*(E_F-E_n^\perp)}}{\hbar},\qquad
 \lambda_n=\frac{2\pi}{k_n},\qquad v_n=\frac{\hbar k_n}{m^*}.
 \label{eq:feasibility_modes}
\end{equation}
The resulting longitudinal wavelengths and velocities appear in
Table~\ref{tab:feasibility_modes}. Counting both spin copies, the ideal
lead density is
\begin{equation}
 n_{1\mathrm D}=\frac{2}{\pi}\sum_{n=1}^3k_n
 =7.94\times10^5\,\mathrm{cm}^{-1}.
 \label{eq:feasibility_density}
\end{equation}
Dividing by $\pi a_ha_v$ gives a cross-section average of
$5.27\times10^{15}\,\mathrm{cm}^{-3}$. This is an occupancy-based estimate;
it is not a uniform doping prescription or a measured Hall density.

Selective-area growth provides experimental precedent for connected InAs
channels and coherent loop transport~\cite{FeasLee2019}.
It does not reproduce the assumed elliptical confinement automatically.
The fabricated cross section, surface electrostatics, dielectric coatings,
and metal gates must be included in a self-consistent confinement model.
This is particularly important for InAs, whose surfaces are known to host
an electron accumulation layer~\cite{Noguchi1991}: without sufficient gate
control, the carrier distribution and transverse spectrum would be set by
this surface charge rather than by the hard-wall profile assumed here.
Subband spectroscopy should establish the operating density. Matching the
two nominal bend outlines is insufficient if their potentials or contacts
differ. Three open lead modes also do not imply three well-transmitting
modes through the split sections, as the scattering results already show.

\subsection{Temperature and phase coherence}
\label{sec:temperature}

We consider an electron temperature of $T_e=100\,\mathrm{mK}$ as an
experimental target. At this temperature,
$k_BT_e=8.62\,\mu\mathrm{eV}$, and the full width at half maximum of
$-\partial f/\partial E$ is approximately $30.4\,\mu\mathrm{eV}$.
The continuum fourth-mode threshold lies about $0.792\,\mathrm{meV}$
above $E_F$, so thermal population of that mode is negligible within the
assumed confinement. The single open arm mode lies $0.56\,\mathrm{meV}$
below $E_F$, about eighteen thermal widths, and remains open throughout
the thermal window.

Define the ballistic thermal length explicitly as
\begin{equation}
 L_{T,n}=\frac{\hbar v_n}{k_BT_e}.
 \label{eq:feasibility_thermal_length}
\end{equation}
The three values at $100\,\mathrm{mK}$ are $21.6$, $15.7$, and
$10.7\,\mu\mathrm m$. The corresponding reference transit times
$L_{\rm ref}/v_n$ are $3.90$, $5.35$, and $7.84\,\mathrm{ps}$.
These estimates support a low-temperature starting point, but neither
the reference transit time nor $L_{T,n}$ accounts for extra dwell time
near a constriction or resonance. Quantitative finite-temperature contrast
requires the energy average
\begin{equation}
 G(T_e)=\frac{2e^2}{h}\int dE\,
 \mathcal T(E)\left(-\frac{\partial f}{\partial E}\right).
 \label{eq:feasibility_thermal_average}
\end{equation}
Evaluating Eq.~\eqref{eq:feasibility_thermal_average} with the energy
dependence of Fig.~\ref{fig:energy}, an electron temperature of
$100\,\mathrm{mK}$ retains $97.7\%$ of the zero-temperature visibility on
the $6\,\mathrm{nm}$ grid and $97.6\%$ on the $8\,\mathrm{nm}$ grid; the
corresponding values are $99.4\%$ at $50\,\mathrm{mK}$ and
$91.0$--$91.4\%$ at $200\,\mathrm{mK}$. Two independent evaluations of
the integral, a spline quadrature and a polynomial fit integrated with
the exact moments of $-\partial f/\partial E$, agree to within
$4\times10^{-6}$ in transmission at $100\,\mathrm{mK}$. The retained
fraction is insensitive to the grid even though the visibility itself is
not, consistent with the grids differing mainly by a rigid energy shift.
The common-mode symmetry survives this
average under the assumptions of Sec.~\ref{sec:symmetry}.

Phase coherence is a separate requirement: the dephasing time must exceed
the dwell times of the interfering contributions. Lee \textit{et al.}
reported an AB-based coherence length of approximately $2\,\mu\mathrm m$
at $50\,\mathrm{mK}$ in selectively grown InAs loops, using a diffusive
interpretation of the temperature dependence~\cite{FeasLee2019}.
This demonstrates a relevant coherence scale, but does not determine the
coherence length of the present few-mode device at $100\,\mathrm{mK}$.

\subsection{Elastic scattering and mobility}

Mobility cannot be obtained from the dimensions and effective mass alone.
A useful conditional estimate follows from a mode-resolved transport time:
\begin{equation}
 \mu_n=\frac{e\tau_{{\rm tr},n}}{m^*},\qquad
 \ell_{{\rm tr},n}=v_n\tau_{{\rm tr},n}
 =\frac{\hbar k_n}{e}\mu_n.
 \label{eq:feasibility_mobility}
\end{equation}
This relaxation-time estimate is a cleanliness benchmark, not a replacement
for a multichannel disorder calculation. Setting
$\ell_{{\rm tr},n}=L_{\rm ref}$ defines
\begin{equation}
 \mu_n^*=\frac{eL_{\rm ref}}{\hbar k_n}.
 \label{eq:feasibility_mobility_target}
\end{equation}
The resulting values are approximately
$(2.98,4.09,6.00)\times10^5\,\mathrm{cm^2\,V^{-1}\,s^{-1}}$.
They mark a mean free path comparable to the reference length; reproducing
the clean transmission curves would generally require
$\ell_{{\rm tr},n}\gg L_{\rm ref}$ for the relevant modes.

\begin{table}[tb]
\centering
\caption{Continuum lead estimates at the chosen Fermi energy.
Velocities are in $10^5\,\mathrm{m\,s^{-1}}$, and $\mu_n^*$ is in
$10^5\,\mathrm{cm^2\,V^{-1}\,s^{-1}}$.
Thermal lengths use $T_e=100\,\mathrm{mK}$.
The mobility column is a conditional target, not a material measurement.}
\label{tab:feasibility_modes}
\begin{tabular}{ccccc}
\toprule
$n$ & $\lambda_n\;(\mathrm{nm})$ & $v_n$ &
$L_{T,n}\;(\mu\mathrm m)$ & $\mu_n^*$\\
\midrule
1 & 112.1 & 2.82 & 21.6 & 2.98\\
2 & 153.9 & 2.05 & 15.7 & 4.09\\
3 & 225.6 & 1.40 & 10.7 & 6.00\\
\bottomrule
\end{tabular}
\end{table}

For comparison, optimized InAs/InGaAs selective-area structures have
shown field-effect mobilities above
$10^4\,\mathrm{cm^2\,V^{-1}\,s^{-1}}$~\cite{FeasBeznasyuk2022}.
If that mobility were assigned to each of the present lead modes,
Eq.~\eqref{eq:feasibility_mobility} would give mean free paths of only
$37$, $27$, and $18\,\mathrm{nm}$. This substitution illustrates the
scale of the challenge; it does not transfer the measured scattering
times between devices with different densities and confinement.
The earlier InAs platform also showed diffusive channels with reported
elastic mean free paths of $20$--$100\,\mathrm{nm}$~\cite{FeasLee2019}.
Subsequent work with InGaAs buffer and capping layers also reports
improvements in mobility and phase coherence in selectively grown InAs
nanowires~\cite{Adhikari2026}. These relative improvements do not directly
determine the elastic mean free path of the present few-mode geometry.

A diffusive sample can remain phase coherent, and exactly symmetric
elastic disorder preserves the stationarity theorem. Neither fact
ensures the clean numerical response. Conversely, Sec.~\ref{sec:robustness}
does not establish a universal mobility threshold for protection:
its disorder amplitude has not been calibrated to a transport mean free
path.

\subsection{Flux magnitude and source dimensions}

The normal-state AB phase uses the flux period $\Phi_0=h/e$:
\begin{equation}
 \phi=2\pi\frac{\Phi}{\Phi_0},\qquad
 \Phi_0=4.13567\times10^{-15}\,\mathrm{Wb}.
 \label{eq:feasibility_flux_quantum}
\end{equation}
The bias $\phi=\pi/4$ used in Sec.~\ref{sec:robustness} requires
\begin{equation}
 \Phi_{\pi/4}=\frac{h}{8e}
 =5.1696\times10^{-16}\,\mathrm{Wb}
 \label{eq:feasibility_target_flux}
\end{equation}
in each opening, with opposite signs for the CW configuration.

The opening short semiaxis, $a_o=17.5\,\mathrm{nm}$, is not the magnetic
core radius. For a centered circular source, the core, insulation,
winding, and placement allowance must fit inside that radius:
\begin{equation}
 r_c+t_{\rm ins}+t_{\rm wind}+\delta_{\rm place}<a_o.
 \label{eq:feasibility_clearance}
\end{equation}
Here the thicknesses denote the total radial allocation to the specified
components. A $35\,\mathrm{nm}$-diameter core leaves no such allowance.

In a long-source estimate with approximately uniform core induction,
$\Phi\simeq B_c\pi r_c^2$. For $\phi=\pi/4$, core radii of
$15$, $12$, and $10\,\mathrm{nm}$ require approximately
$0.731$, $1.143$, and $1.646\,\mathrm T$, respectively.
The flux-carrying core area, rather than the whole elliptical opening,
enters this estimate. If $B_c=2.15\,\mathrm T$ is adopted as an
illustrative induction, the required radius is $r_c\simeq8.75\,\mathrm{nm}$.
This induction is essentially the saturation value of iron,
$\mu_0M_s\simeq2.15\,\mathrm T$, so the estimate leaves no margin in the
core material.
This is a flux-budget calculation, not a prediction of the remanent
induction of an individual nanowire.

Using the same induction with $r_c=17.5\,\mathrm{nm}$ instead gives
$\phi\simeq1.00034\pi$. This is not the $\pi/4$ operating point.
Moreover, at exactly $\phi=\pi$, the assignments $(\pi,-\pi)$ and
$(\pi,\pi)$ are equivalent modulo the flux periods of the ideal model.
A comparison of CW and SW common-mode sensitivity must therefore use a
bias away from integer multiples of $\pi$.

Iron nanowire arrays with $35\,\mathrm{nm}$ diameter have shown strong
axial remanence and coercivity~\cite{FeasSun2007}. Those array measurements
do not establish the properties of a smaller isolated core embedded
between semiconductor arms. Core oxidation, domain structure, switching,
and the local magnetic environment must be characterized. A remanent
core and a current-driven solenoid are also different implementations.
For an illustrative $2\,\mu\mathrm m$ air-core solenoid of radius
$15\,\mathrm{nm}$, $B\simeq\mu_0NI/L_s$ gives
$NI\simeq1.16\,\mathrm{A\!\cdot turn}$ at the target flux.
Distributed over the $2\,\mu\mathrm m$ length at a $10\,\mathrm{nm}$
winding pitch, this is about $6\,\mathrm{mA}$ per turn, or a current
density of order $6\times10^9\,\mathrm{A\,cm^{-2}}$ in a
$10\times10\,\mathrm{nm}^2$ conductor, far above the
$10^6$--$10^7\,\mathrm{A\,cm^{-2}}$ typical of metallic interconnects.
A ferromagnetic core changes this relation through its nonlinear
magnetization. Turn count and current cannot be selected without the
winding dimensions, resistance, current-density limit, and cryogenic
heat budget $P=I^2R_{\rm coil}$. Because the required flux $h/8e$ does
not depend on the aperture, a larger opening would relax the clearance
condition~\eqref{eq:feasibility_clearance}; it would also change the
transport problem and require a separate calculation.

\subsection{Leakage fields and spin-dependent terms}
\label{sec:leakage}

A finite solenoid or magnetic core has return fields. Zero externally
applied field does not imply $\mathbf B=0$ inside the conductor.
For orientation only, a uniformly magnetized slender cylinder of length
$L_s$, core induction $B_c\simeq\mu_0M$, and radius $r_c$ has an exterior
midplane return-field scale
\begin{equation}
 |B_{{\rm ret},z}|\simeq\frac{2B_cr_c^2}{L_s^2},
 \qquad r_c<r_\perp\ll L_s/2.
 \label{eq:feasibility_return_field}
\end{equation}
For the illustrative $r_c=8.75\,\mathrm{nm}$,
$B_c=2.15\,\mathrm T$, and $L_s=2\,\mu\mathrm m$ source, this scale is
$0.082\,\mathrm{mT}$. It is neither a bound over the device nor a
calculation for the two-source assembly. A magnetostatic solution must
include both sources, their ends, and any return or shielding structure,
followed by transport using the resulting field in the conducting volume.

Field exclusion is a physical design requirement: the shielded AB
experiment of Tonomura \textit{et al.} used a coated toroidal magnet to
separate the magnetic flux from the electron wave~\cite{FeasTonomura1986}.
That result does not provide a fabrication solution for the present
$35\,\mathrm{nm}$ aperture.

InAs also requires a spinful assessment, and its spin--orbit coupling is
not weak on the scale of the device. The preceding calculations do not
include Rashba or Dresselhaus coupling, Zeeman splitting, or a
source-induced electrostatic potential. Rashba coefficients of
$0.06$--$0.2\,\mathrm{eV}\,\text{\AA}$ have been reported in gated InAs
nanowires~\cite{Takase2017}. For $\alpha=0.1$--$0.2\,\mathrm{eV}\,\text{\AA}$
the spin--orbit energy $\alpha k_0$ is $0.7$--$1.3\,\mathrm{meV}$, and
the spin--orbit length $\ell_{\rm so}=\hbar^2/(m^*\alpha)\simeq
170$--$330\,\mathrm{nm}$ is shorter than $L_{\rm ref}$. The spin then
precesses by roughly $7$--$13\,$rad along the guide, and the modulation
amplitudes and response coefficients of
Secs.~\ref{sec:decoherence_analysis} and~\ref{sec:robustness}, obtained
without spin--orbit coupling, cannot be carried over to such a device.
The symmetry statement is less sensitive. By Sec.~\ref{sec:symmetry}, a
Rashba coupling that is equal in the two modules, lateral spin--orbit
terms of a $C_2$-invariant confinement, and the Dresselhaus term for a
$[001]$ orientation all preserve Eq.~\eqref{eq:full_composite_condition},
whereas a difference between the gate fields of the two modules acts
like $V_-$. The same holds for the return fields: those of two identical
sources in $C_2$-related positions transform with the source fluxes and
leave Eq.~\eqref{eq:full_composite_condition} intact, both orbitally and
through the Zeeman term, while a difference between the sources acts
like a flux mismatch. The Zeeman energy at the $0.082\,\mathrm{mT}$
scale above is small in any case: even with the bulk factor
$|g^*|\simeq14.7$~\cite{Gawarecki2020}, $|g^*|\mu_BB\simeq0.07\,
\mu\mathrm{eV}$, two orders of magnitude below $k_BT_e$ at
$100\,\mathrm{mK}$. A quantitative spinful calculation for a specified
crystal orientation and gate layout remains outside the present work.

\subsection{Readout and a test of stationarity}
\label{sec:readout}

The finest-grid zero-temperature sweep gives
$G\simeq127.7$--$137.1\,\mu\mathrm S$ and
$R_{2\rm t}\simeq7.29$--$7.83\,\mathrm{k}\Omega$.
These are Landauer two-terminal values, including the quantum contact
contribution, rather than a bulk resistivity inferred from mobility.
At an illustrative bias of $V_{\rm sd}=1\,\mu\mathrm V$, the full sweep
corresponds to a current modulation of approximately $9.42\,\mathrm{pA}$.

Detecting stationarity is more demanding than resolving that full swing.
At $\phi=\pi/4$, the local clean-model estimate is
\begin{equation}
 \Delta I_{\rm CW}\simeq
 V_{\rm sd}\frac{2e^2}{h}\frac{K_c}{2}c^2.
 \label{eq:feasibility_readout}
\end{equation}
With $K_c=-0.1197\,\mathrm{rad}^{-2}$, a common offset
$c=0.1\,\mathrm{rad}$ gives
$|\Delta I_{\rm CW}|\simeq0.046\,\mathrm{pA}$
at the stated voltage. The corresponding SW linear estimate is
$0.55\,\mathrm{pA}$. Both estimates inherit the local-expansion and
finite-grid limitations. The energy-referenced curvature of
Sec.~\ref{sec:cw_sweep} would reduce the counter-wound estimate to
$0.036$--$0.039\,\mathrm{pA}$, and both values depend on the operating
energy (Sec.~\ref{sec:energy}). For comparison, the equilibrium current-noise
scale $\sqrt{4k_BT_eG}$ is about
$0.028\,\mathrm{pA}/\sqrt{\mathrm{Hz}}$ at $100\,\mathrm{mK}$. Against
this scale alone, a $0.04\,\mathrm{pA}$ signal reaches a signal-to-noise
ratio of ten after roughly a minute of averaging.
This excludes amplifier noise and low-frequency drift and is not an
experimental signal-to-noise prediction.

A direct test should calibrate the two linked fluxes, hold the
differential coordinate fixed, and compare the conductance at positive
and negative common offsets. Equal changes of two control currents do
not necessarily produce equal changes of phase. The two kinds of flux
noise must also be distinguished. A common current fluctuation through
oppositely wound coils generally produces a differential flux
fluctuation, which is not the protected direction for CW bias, so
cross-coupling between the controls must be measured. A drift of the
perpendicular ambient field, by contrast, enters both openings with a
common sign and is stationary to first order by Eq.~\eqref{eq:ambient},
including its part inside the semiconductor. A uniform perpendicular
applied field therefore provides
a complementary test that requires no calibration of the two sources
against each other: at fixed source fluxes and counter-wound bias the
conductance must be even in the applied field, whereas at same-winding
bias it responds linearly (Sec.~\ref{sec:ambient}).

Existing InAs transport and magnetic-nanowire experiments support
individual ingredients of the proposal. They do not demonstrate their
integration in the specified geometry. The immediate experimental
requirements are measured few-mode confinement and coherence, a scalar
potential matched between the modules, and a flux-source design with
adequate clearance and quantified leakage. Until these are established,
the present device is a conditional experimental proposal rather than
a demonstrated implementation of the ideal response.

\section{Discussion}
\label{sec:discussion}

The principal result is a constraint on the transmission of a balanced
three-dimensional scattering structure. Opposite fluxes alone do not
make the device insensitive to flux. As shown in
Sec.~\ref{sec:phases}, their magnetic contributions cancel for contours
with equal winding numbers about the two openings, whereas other
contours retain a flux-dependent phase. The oscillations in
Fig.~\ref{fig:CW_transmission} are consistent with this distinction.
Recombination between the bends does not preserve a unique pair of
arms through the entire device, and the reference length identity in
Eq.~\eqref{eq:pathbalance} does not establish equality of all propagation
phases.

The transmission symmetry instead follows from microreversibility,
two-terminal unitarity, and the rotation that exchanges the complete
modules and their contacts. In the coordinates of
Eq.~\eqref{eq:total_flux_coordinates}, it requires
$F_E(C,D)=F_E(-C,D)=F_E(C,-D)$. Consequently, $C=0$ is a line of
stationarity with respect to common offsets wherever the derivative
exists. Its location follows from symmetry, while the curvature of the
response depends on confinement, energy, and scattering. Earlier
analyses of two flux vortices concern free-space
scattering~\cite{SymmetryBogomolny2010,SymmetryBogomolny2016}.
Here the observable is the total transmission between reservoirs, and
the two independently controlled fluxes specify the direction in which
its response is stationary. Any operation that exchanges the complete
modules and contacts would serve equally; the opposite curvature of the
present geometry is a matter of layout (Sec.~\ref{sec:symmetry}).

This distinction also sets the comparison with the same-winding
configuration. Counter-wound and same-winding biases suppress the
linear response to common and differential offsets, respectively.
Their usefulness therefore depends on the fluctuations present in a
sample. A differential fluctuation generally changes the counter-wound
transmission to first order even though common-mode stationarity
persists at the shifted differential bias. The symmetry does not
suppress both directions of noise, and it does not establish an ordering
of the two conductances. Biases at $\phi=m\pi$ are especially unsuitable
for distinguishing the assignments, since they are equivalent modulo
the individual flux periods.

The size of the protected perturbation also depends on the device scale.
For the present dimensions, the linear same-winding response to a uniform
perpendicular field found at $\phi=\pi/4$ on the $8\,\mathrm{nm}$ grid in
Sec.~\ref{sec:ambient}, $2.2\times10^{-3}$ per $\mathrm{mT}$, corresponds
to a change of order $10^{-6}$ for a $1\,\mu\mathrm T$ drift.
Common-mode stationarity is therefore most relevant for correlated
offsets of the flux controls, for example of two identically wound,
separately driven sources, and for larger devices, whose coupling to a
uniform field grows with the enclosed area. Fluctuations of a symmetric
scalar potential change the transmission directly and are not addressed
by the flux symmetry.

The counter-wound assignment has a direct analogue in gradiometric
superconducting loops, in which a figure-eight winding makes the net flux
of a uniform field vanish; such designs are standard in SQUID
magnetometry~\cite{clarke2004squid}, and a recent preprint reports
improved coherence and frequency stability in a gradiometric
transmon~\cite{fu2026fluxnoise}. In a gradiometer a single winding
pattern cancels the uniform-field flux exactly. Here recombination and
reflection generate contours with all combinations of winding numbers
(Sec.~\ref{sec:phases}), and a common offset enters many of them. What
is removed is only the odd part of the total transmission, by the
combined action of microreversibility and the module exchange. The
protection is therefore first order, but it does not rely on balancing
individual contours: it holds for any number of channels and any amount
of elastic reflection, along the entire line $C=0$, and extends to a
uniform field that also penetrates the conductor
(Sec.~\ref{sec:ambient}). The present calculation concerns
normal-state transmission and does not predict a superconducting-qubit
coherence time.

Flux-noise correlations have also been measured in a two-loop
superconducting qubit~\cite{gustavsson2011}. This illustrates why
common and differential fluctuations need not have equal statistical
weight. Those measurements do not determine the flux noise of the
proposed InAs device or the scalar-disorder correlation $\rho$ used
in Sec.~\ref{sec:robustness}.

Section~\ref{sec:decoherence_analysis} makes a further distinction
between the mean signal and its fluctuations. For purely common,
quasistatic Gaussian noise, the leading transmission variance at a
balanced counter-wound bias is proportional to $\sigma_c^4$; a reference
with nonzero slope instead has a contribution proportional to
$\sigma_c^2$. The mean transmission shift remains quadratic in both
cases. At $\phi=\pi/4$, the calculated normalized mean-shift coefficient
is larger in magnitude for the counter-wound configuration, despite its
suppressed leading variance. Reduced fluctuations therefore need not
mean better retention of the mean transmission. These statements concern
an ensemble of coherent static problems and do not establish a longer
phase-coherence time.

The three-dimensional calculation retains transverse confinement,
reflection, and mode conversion around the openings. Although three
orbital channels propagate in the leads, the third incident mode is
strongly reflected by the constrictions. Neither equal splitting nor
single-mode propagation is needed for the total-transmission identity.
Its applicability to an injected electron beam nevertheless depends on
how that beam is prepared and detected: the summed reservoir
transmission generally does not impose the same parity on an individual
mode or an arbitrarily prepared coherent superposition.

The numerical results establish a finite-grid reference for this
geometry and identify the origin of its grid dependence. The finest
sampled sweep has a visibility of approximately $3.56\%$. Most of the
change between meshes at fixed $E_F$ comes from the discretization shift
of the single open arm mode, which moves the whole energy structure of
the response; referring the grids to a common threshold gives an
estimated continuum visibility of $2.6$--$2.9\%$. The magnitudes left
free by the symmetry vary on a scale of $0.1\,\mathrm{meV}$, so the
operating energy must be specified, and ideally controlled, to a few
tens of $\mu\mathrm{eV}$ if the quoted tolerances are to apply; finite
temperature, by contrast, reduces the visibility only by about $2\%$ at
$100\,\mathrm{mK}$. The small symmetry residual verifies a property of the discretized scattering problem; it does not provide an equally small uncertainty in the continuum modulation amplitude.

Section~\ref{sec:robustness} identifies which imperfections restore a
linear common-mode response. Symmetric scalar disorder preserves
stationarity, even when it reduces transmission. The antisymmetric
part of a weak potential produces the leading slope. The result
$s_{\rm rms}\propto W\sqrt{1-\rho}$ follows from the specified spatial
covariance and the calculated response kernel; it is not a universal
law relating mobility to protection. Likewise, unequal flux magnitudes
can be compensated by shifting the common bias in an otherwise balanced
structure, whereas that correction does not generally repair an
asymmetric confining potential.

The feasibility estimates in Sec.~\ref{sec:feasibility} leave two separate
experimental requirements: coherent transport through the patterned
InAs guide and controlled fluxes with sufficiently small fields in the
conducting region. The opening diameter $35\,\mathrm{nm}$ is a space
budget for the entire source assembly. It is not available core
diameter once insulation and windings are included. Spin-dependent
terms and leakage fields must also satisfy the full covariance
condition in Eq.~\eqref{eq:full_composite_condition} before the orbital
symmetry argument can be extended to a fabricated device.

A direct test would resolve the odd part of the common-offset response
at fixed differential bias,
\begin{equation}
\mathcal T_{\rm odd}(c;D)
=\frac{F_{E_F}(c,D)-F_{E_F}(-c,D)}{2}.
\label{eq:discussion_odd_response}
\end{equation}
It vanishes in the balanced model. Measuring this quantity over several
values of $D$, then introducing a controlled electrostatic imbalance,
would distinguish the predicted symmetry from an isolated accidental
extremum. Such a test requires calibration of the two flux controls:
varying a shared current through oppositely wound sources generally
changes $D$, rather than scanning the common coordinate $C$. A uniform
perpendicular field offers a second test of the same symmetry, as described in
Sec.~\ref{sec:readout}.

\section{Scope and limitations}
\label{sec:scope}

The symmetry result concerns the total elastic transmission of a
two-terminal conductor with the Hamiltonian and boundary conditions
specified in Sec.~\ref{sec:model}. It requires covariance of the complete
scattering structure, including the openings, scalar potentials, and
contacts. At a counter-wound bias it constrains the response to a common
flux offset, wherever that response is differentiable. It does not
require reflectionless propagation, but neither does it guarantee a
large conductance, high visibility, or transmission of a particular
incident mode. It is therefore insufficient, on its own, to establish
the fidelity of an injected electron state.

The electronic calculation uses a spin-independent, parabolic
effective-mass Hamiltonian with hard-wall confinement. Spin enters only
as a degeneracy factor. Spin--orbit coupling, Zeeman splitting, and
self-consistent electrostatics are not included. In a physical InAs
structure, surface charge, gate fields, and the confining interfaces
would determine the carrier distribution and transverse spectrum.
The chosen Fermi energy and channel occupation are model parameters,
not measured properties of a fabricated device; because the response
magnitudes vary on a $0.1\,\mathrm{meV}$ scale (Sec.~\ref{sec:energy}),
they are specific to that choice. Any spinful extension
must satisfy the full transformation condition in
Eq.~\eqref{eq:full_composite_condition}; time-reversal invariance of an
individual spin--orbit term is not sufficient. For Rashba coupling
typical of gated InAs the spin--orbit length is shorter than
$L_{\rm ref}$ (Sec.~\ref{sec:leakage}), so the numerical magnitudes
reported here would change in such a device even where the symmetry
survives.

The magnetic model assumes flux confined to the excluded openings and
zero magnetic field throughout the conductor. A spatially uniform
perpendicular field is compatible with the protection (Sec.~\ref{sec:ambient}), and so are
the return fields of two identical sources in $C_2$-related positions
(Sec.~\ref{sec:leakage}); an ambient field gradient or a difference
between the sources is not. The source estimates in
Sec.~\ref{sec:feasibility} do not establish a field-free conductor for a
finite iron-core solenoid. A realizable assembly requires a magnetostatic
calculation of both sources, their return fields, and their coupling to
the controls. The resulting vector potential and any field-dependent
terms must then be included in transport. Fitting a source within the
opening and obtaining the desired enclosed flux are necessary geometric
and magnetic checks, but do not by themselves validate the field-free
Hamiltonian.

The noise calculations describe quasistatic flux offsets and ensembles
of static scalar potentials. They contain no noise spectrum, inelastic
scattering process, or microscopic dephasing mechanism. The quartic
small-noise transmission variance applies to a purely common Gaussian
offset in the balanced structure. Differential noise or a residual
common-mode slope can restore a contribution quadratic in noise
amplitude. No dephasing time or phase-coherence length is extracted from
these calculations. Thermal averaging preserves the symmetry when its
statistical weight is flux independent; at $100\,\mathrm{mK}$ it reduces
the calculated visibility by about $2\%$.

The numerical values in Secs.~\ref{sec:decoherence_analysis}
and~\ref{sec:robustness} remain finite-grid estimates. The mesh
comparisons show that the grid dependence at fixed energy is dominated
by the discretization shift of the arm threshold. Referring the grids to
a common threshold gives continuum estimates of the visibility and
response coefficients about $15$--$30\%$ smaller in magnitude than their
$3\,\mathrm{nm}$ values; these estimates rest on that referencing and
have not been confirmed on a finer grid. The energy-resolved and
finite-temperature results were obtained on the $8$ and $6\,\mathrm{nm}$
grids. The local drift windows are Taylor estimates. Their finite-offset
accuracy is tested in Sec.~\ref{sec:decoherence_analysis} by a full
common-offset sweep at $\phi=\pi/4$ on the $3\,\mathrm{nm}$ grid.
This does not establish continuum drift tolerances or their validity
at other energies. The disorder scaling uses a weak-potential
expansion with a prescribed covariance. Selected
finite-perturbation checks support these expansions locally; they do
not establish their accuracy for arbitrary noise or disorder strength.

Finally, the adopted opening aspect ratio and transverse dimensions
define one reference design. The calculations do not optimize that
design or determine tolerances for roughness, unequal cross sections,
or displacement of the openings. The disorder amplitude has not been
calibrated to an elastic mean free path, so it cannot be converted into
a measured mobility threshold. The feasibility estimates identify
requirements for testing the model; they do not demonstrate that one
device can simultaneously meet its transport, symmetry, source-size,
and magnetic-confinement assumptions.

\section{Conclusion}
\label{sec:conclusion}

We have formulated a three-dimensional model of an InAs waveguide with
two quarter-circle bends of opposite curvature, elliptical transverse
confinement, and a through-opening in each module that accommodates an
independently specified confined flux. For a balanced two-terminal
structure, microreversibility and the rotation exchanging the modules
require the total transmission to be even in each of the common and
differential flux coordinates. The counter-wound configuration therefore
lies on a line of first-order stationarity against common offsets, and
the same-winding configuration on the corresponding line for
differential offsets. Within the orbital model these identities hold
independently of the number of propagating channels and the amount of
elastic reflection. The protected direction
includes a uniform drift of the perpendicular ambient field, even though
that field penetrates the conductor, and the opposite curvature of the
bends is a choice of layout rather than a requirement of the symmetry.

Three-dimensional scattering calculations at $R=350\,\mathrm{nm}$ and
$E_F\simeq7.283\,\mathrm{meV}$ show how this constraint coexists with
transverse confinement and strong reflection at the openings. The finest
sampled counter-wound sweep has a visibility of approximately $3.56\%$;
its grid dependence is dominated by a discretization shift of the single
open arm mode, and referring the grids to a common threshold gives an
estimated continuum visibility of $2.6$--$2.9\%$. The response
magnitudes vary on a $0.1\,\mathrm{meV}$ scale, whereas a
$100\,\mathrm{mK}$ electron temperature retains about $98\%$ of the
visibility. Under purely common, weak Gaussian phase noise the leading
transmission variance at a counter-wound bias is quartic in the noise
amplitude, while the mean shift remains quadratic. A flux mismatch can be compensated by
recentering the common bias in an otherwise symmetric structure, whereas
the antisymmetric part of a scalar potential restores a linear response,
with a weak-disorder sensitivity that scales as $W\sqrt{1-\rho}$ for the
prescribed covariance.

An experimental test requires coherent few-mode transport, calibrated
control of the two fluxes, and a full Hamiltonian, including leakage
fields and spin-dependent terms, that satisfies
Eq.~\eqref{eq:full_composite_condition}. Measuring the odd part of the
common-offset response across several counter-wound biases, before and
after introducing a controlled electrostatic imbalance, would test the
predicted mechanism directly; a uniform perpendicular field provides a
complementary test that does not require the two sources to be
calibrated against each other.

\section*{Data availability}
The data that support the findings of this article are not publicly
available. The data, together with the code of the independent
implementation described in Sec.~\ref{sec:decoherence_analysis}, are
available from the authors upon reasonable request.

\begin{acknowledgments}
The authors thank Prof.~Bora Isildak for reviewing an early draft of the
manuscript and for offering helpful comments and suggestions during the
initial development of the work.

The authors used Claude (Anthropic) for language
editing of the manuscript and for assistance in writing and running the
code that produced the energy-resolved, finite-temperature,
grid-alignment, grid-referenced, and uniform-field results, including
Fig.~\ref{fig:energy}. That code was checked against the authors' independent calculations of
the counter-wound sweeps in Fig.~\ref{fig:CW_transmission} and of the entries in
Tables~\ref{tab:response_mesh} and~\ref{tab:robustness_mesh}. All scientific content, derivations, and conclusions were
reviewed by the authors, who take full responsibility for the article.
\end{acknowledgments}

\bibliography{References}
\end{document}